%% file: main.tex
\documentclass[sigconf]{acmart}
\renewcommand\footnotetextcopyrightpermission[1]{}
\usepackage{times}
\usepackage[utf8]{inputenc}
\usepackage{url}
\usepackage{amsmath}
\usepackage{array,multirow}
\usepackage{xspace}
\usepackage{xcolor}
\usepackage{color, colortbl} 
\usepackage{balance}
\usepackage{paralist}
\usepackage{graphicx}
\usepackage{pifont}
\usepackage{array}
\usepackage{booktabs}
\usepackage{tikz}
\usepackage{bm}
\usepackage{enumitem}
\usepackage{todonotes}
\usepackage{arydshln}
\usepackage{makecell}
\usepackage{tabularx}
\usepackage{caption}
\usepackage{soul,color}
\usepackage{mwe}
\usepackage{ifthen}
\usepackage[noend]{algpseudocode}
\usepackage{epsfig}
\usepackage{algorithm}
\usepackage{algpseudocode}
\usepackage{caption}
\usepackage{subcaption}
\usepackage{amsthm}

\usepackage{bbm}
\usepackage{mathtools}

\usepackage{cleveref}

\AtEndPreamble{
  \pagenumbering{arabic}
  \RequirePackage[unicode]{hyperref}
  \hypersetup{colorlinks=true, linkcolor=blue, urlcolor=blue, citecolor=blue}
}

\makeatletter
\renewcommand{\ALG@beginalgorithmic}{\small}

\makeatother

\begin{document}
\input{macros}


\title{PANDAS: Peer-to-peer, Adaptive Networking for Data Availability Sampling within Ethereum Consensus Timebounds}

\author{Matthieu Pigaglio}
\affiliation{%
    \institution{UCLouvain}
    \country{Belgium}
}

\author{Onur Ascigil}
\affiliation{%
    \institution{Lancaster University}
    \country{United Kingdom}
}

\author{Michał Król}
\affiliation{%
    \institution{City, University of London}
    \country{United Kingdom}
}

\author{Sergi Rene}
\affiliation{%
    \institution{Datahop Labs}
    \country{United Kingdom}
}

\author{Felix Lange}
\affiliation{%
    \institution{Ethereum Foundation}
    \country{Germany}
}

\author{Kaleem Peeroo}
\affiliation{%
    \institution{City, University of London}
    \country{United Kingdom}
}

\author{Ramin Sadre}
\affiliation{%
    \institution{UCLouvain}
    \country{Belgium}
}

\author{Vladimir Stankovic}
\affiliation{%
    \institution{City, University of London}
    \country{United Kingdom}
}

\author{Etienne Rivière}
\affiliation{%
    \institution{UCLouvain}
    \country{Belgium}
}
\email{etienne.riviere@uclouvain.be}


\input{sections/abstract}


\maketitle
\pagestyle{plain}

\input{sections/intro}

\input{sections/background}

\input{sections/das}

\input{sections/overview}
\input{sections/assignment}
\input{sections/pandas_phases}
\input{sections/fetching}
\input{sections/evaluation}

\input{sections/discussion}
\input{sections/related}
\input{sections/conclusion}

\begin{acks}
This work was supported by a grant from the Ethereum foundation (grant \#FY22-0753).
Ramin Sadre and Etienne Rivière are supported by Belgian Wallonia region CyberExcellence program (grant \#2110186).
\end{acks}

\bibliographystyle{plain}
\bibliography{refs}


\end{document}

%% file: macros.tex


\newboolean{showcomments}
\setboolean{showcomments}{false}

\newcommand{\vspacebeforesection}{\vspace{-0mm}}
\newcommand{\vspaceaftersection}{\vspace{-0mm}}
\newcommand{\vspacebeforesubsection}{\vspace{-0mm}}
\newcommand{\vspaceaftersubsection}{\vspace{-0mm}}

\newcommand{\vspacebeforecaption}{\vspace{-0mm}}
\newcommand{\vspaceaftercaption}{\vspace{-0mm}}


\ifthenelse{\boolean{showcomments}}
{ \newcommand{\mynote}[3]{
    \protect\fbox{\bfseries\sffamily\scriptsize#1}
    {\small\textsf{\emph{\color{#3}{#2}}}}}}
{ \newcommand{\mynote}[3]{}}

\newcommand{\er}[1]{\mynote{Etienne}{#1}{blue}}
\newcommand{\michal}[1]{\mynote{Michał}{#1}{brown}}
\newcommand{\sergi}[1]{\mynote{Sergi}{#1}{violet}}
\newcommand{\onur}[1]{\mynote{Onur}{#1}{red}}
\newcommand{\kp}[1]{\mynote{Kaleem}{#1}{orange}}
\newcommand{\rs}[1]{\mynote{Ramin}{#1}{green}}
\newcommand{\ef}[1]{\mynote{EF}{#1}{green}}
\newcommand{\mat}[1]{\mynote{Matthieu}{#1}{orange}}

\newcommand\sysname{PANDAS\xspace}


\newcommand\attacker{attacker\xspace}
\newcommand\adversary{attacker\xspace}
\newcommand\prerecorded{prerecorded\xspace}
\newcommand\prerecords{prerecords\xspace}
\newcommand\prerecord{prerecord\xspace}
\newcommand\bitswap{\emph{Bitswap}\xspace}

\newcommand{\cf}{cf.\@\xspace}
\newcommand{\vs}{vs.\@\xspace}
\newcommand{\etc}{etc.\@\xspace}
\newcommand{\ala}{ala\@\xspace}
\newcommand{\wrt}{w.r.t.\@\xspace}
\newcommand{\etal}{\textit{et al.}\@\xspace}
\newcommand{\eg}{e.g.,\@\xspace}
\newcommand{\Eg}{E.g.,\@\xspace}
\newcommand{\ie}{i.e.,\@\xspace}
\newcommand{\Ie}{I.e.,\@\xspace}
\newcommand{\via}{\textit{via}\@\xspace}
\newcommand{\defacto}{\textit{de facto}\@\xspace}

\newcommand\mypara[1]{\vspace{0.05in} \noindent \textbf{#1.}}
\newcommand\para[1]{\vspace{0.05in} \noindent \textbf{#1.}}


\def\sampler{sampler\xspace}

\newcommand\inlinesection[1]{{\bf #1.}}

\def\first{({\it i})\xspace}
\def\second{({\it ii})\xspace}
\def\third{({\it iii})\xspace}
\def\fourth{({\it iv})\xspace}
\def\fifth{({\it v})\xspace}
\def\sixth{({\it vi})\xspace}

\newcommand{\one}{({\em i})\xspace}
\newcommand{\two}{({\em ii})\xspace}
\newcommand{\three}{({\em iii})\xspace}
\newcommand{\four}{({\em iv})\xspace}
\newcommand{\five}{({\em v})\xspace}

\definecolor{verylightgray}{gray}{0.8}

\newcolumntype{L}{l<{\hspace{1cm}}}
\newcolumntype{C}{c<{\hspace{1cm}}}
\newcolumntype{D}{c<{\hspace{0.3cm}}}

\newcommand\vgap{\vskip 2ex}
\newcommand\marker{\vgap\ding{118}\xspace}

\def\na{--}
\def\unsure{?}
\def\missing{$!$}
\newcommand{\yes}{\ding{51}}
\newcommand{\no}{\ding{55}}
\DeclareRobustCommand\pie[1]{
\tikz[every node/.style={inner sep=0,outer sep=0, scale=1.5}]{
\node[minimum size=1.5ex] at (0,-1.5ex) {}; 
 \draw[fill=white] (0,-1.5ex) circle (0.75ex); \draw[fill=black] (0.75ex,-1.5ex) arc (0:#1:0.75ex); 
}
}
\def\L{\pie{0}} 
\def\M{\pie{-180}} 
\def\H{\pie{360}} 

\newcommand{\cmark}{\ding{51}}%
\newcommand{\xmark}{\ding{55}}%

\newcommand{\peerid}{\ensuremath{\mathsf{peerid}}}
\newcommand{\cid}{\ensuremath{\mathsf{cid}}}
\newcommand{\key}{\ensuremath{\mathsf{key}}}
\newcommand{\threshold}{\ensuremath{\mathsf{thr}}}

\newcommand{\numSybils}{\ensuremath{e}}
\newcommand{\cgen}{\ensuremath{c_{\mathrm{gen}}}}
\newcommand{\twarmup}{\ensuremath{t_{\mathrm{w}}}}
\newcommand{\teff}{\ensuremath{t_{\mathrm{eff}}}}
\newcommand{\coper}{\ensuremath{c_{\mathrm{oper}}}}
\newcommand{\catt}{\ensuremath{c_{\mathrm{att}}}}

\newcommand{\algvar}[1]{\ensuremath{\mathsf{#1}}}

\newcommand{\omnifiguresscalingfactor}{0.56}

\newcommand{\plotscale}{0.28}

\newcommand{\where}{\;|\;}

%% file: sections/abstract.tex
\begin{abstract}
Layer-2 protocols can assist Ethereum's limited throughput, but globally broadcasting layer-2 data limits their scalability.
The \emph{Danksharding} evolution of Ethereum aims to support the selective distribution of layer-2 data, whose availability in the network is verified using randomized data availability sampling (DAS).
Integrating DAS into Ethereum’s consensus process is challenging, as pieces of layer-2 data must be disseminated and sampled within four seconds of the beginning of each consensus slot.
No existing solution can support dissemination and sampling under such strict time bounds.

We propose \sysname, a practical approach to integrate DAS with Ethereum under Danksharding's requirements without modifying its protocols for consensus and node discovery.
\sysname disseminates layer-2 data and samples its availability using lightweight, direct exchanges. 
Its design accounts for message loss, node failures, and unresponsive participants while anticipating the need to scale out the Ethereum network.
Our evaluation of \sysname's prototype in a 1,000-node cluster and simulations for up to 20,000 peers shows that it allows layer-2 data dissemination and sampling under planetary-scale latencies within the 4-second deadline. \er{Adapt the abstract to match the final results we have}
\end{abstract}

%% file: sections/intro.tex
\vspacebeforesection
\section{Introduction}
\vspaceaftersection
\label{sec:intro}


Ethereum, the largest blockchain supporting smart contracts, currently supports adding fewer than a few tens of transactions per second to its main (layer-1) chain.
Complementarily to layer-1 scalability improvements~\cite{hafid2020scaling,rebello2024survey}, expanding support for layer-2 protocols~\cite{gangwal2023survey} is now a priority for the Ethereum community~\cite{ethereum2024roadmap}.

Layer-2 protocols such as \emph{side chains} and \emph{rollups} have the potential to process large amounts of transactions~\cite{gangwal2023survey,jourenko2019sok}.
These protocols periodically produce compressed or batched layer-2 transaction data, which they make available via the layer-1 blockchain.
For instance, participants in an optimistic rollup can download this data, verify its correctness, and submit fraud proofs~\cite{ethereum2024optimistic_rollups,kalodner2018arbitrum,optimism}.

The throughput of layer-2 protocols depends on how much data they can attach to layer-1 blocks.
Previously, the only solution was adding layer-2 data as costly \emph{calldata} transactions.
These transactions competed for permanent block space with other layer-1 transactions, such as ETH transfers and DeFi interactions.
In March 2024, the EIP-4844 (\emph{Proto-Dank\-shar\-ding}$^\text{1}$) proposal~\cite{ethereum_protodanksharding_eip4844} introduced the notion of \emph{blobspace}.
Layer-2 data can now be shared as opaque \emph{binary objects} (blobs).
Blobs are broadcast separately and referenced by \emph{blob-carrying} transactions in the block, which include cryptographic commitments to their content.
Nodes participating in consensus verify these commitments and make blob data available to layer-2 participants for a limited time (4,096 epochs, $\sim$18 days).
While improving over \emph{calldata} transactions regarding costs and supported volume, EIP-4844 still requires blob data to be broadcast and received by all nodes.


A significant upcoming step towards scaling support for layer-2 data in Ethereum is implementing \emph{data availability sampling} (DAS).
The Ethereum \emph{Danksharding}\footnote{%
    \emph{Danksharding} is named after Dankrad Feist, an Ethereum researcher. \emph{Proto-danksharding} is named after him and Diederik Loerakker (\emph{protolambda}).}~\cite{ethereum2024roadmap_danksharding} roadmap plans to support up to 32~MB blob data referenced by each layer-1 block.

To avoid broadcasting this volume of data globally, blob data is erasure-coded, split, and distributed as collections of \emph{cells}, so that each node holds only a fraction of the data.
This shift introduces a new challenge: no single node can independently verify the availability of the full blob.
To address this, Ethereum plans to adopt \emph{data availability sampling} (DAS), wherein nodes collect random \emph{sample} cells from the network until they reach overwhelming confidence that the complete data can be reconstructed.
The Danksharding parameters imply sending 140~MB of erasure-coded cells to the network, with each node randomly sampling 73 cells (40~KB).



Integrating DAS and Ethereum consensus is a challenge.
The need for each node to collect randomly selected samples results in a multitude of exchanges over high-latency links.
At the same time, Ethereum consensus imposes tight time constraints.
A new block is generated every 12 seconds.
A committee must validate each new block within the first four seconds after its creation.
To avoid changes to the consensus protocol, DAS must also be completed within four seconds, allowing committee members to attest block validity and blob data availability simultaneously.

Several approaches are under discussion for integrating DAS with Ethereum consensus, e.g., EIP-7594 (PeerDAS)~\cite{eip-7594}, SubnetDAS~\cite{subnetdas}, or FullDAS~\cite{fulldas}.
These leverage peer-to-peer networks already used by other Ethereum functions, particularly GossipSub~\cite{gossipsub}, a broadcasting network that pre-establishes dissemination overlays and uses them as channels for gossiping data.
There is a discrepancy between the high costs of establishing new channels and the randomized nature of DAS.
Multi-hop gossiping also has inherently high latency.
These two factors lead these proposals to suggest that random sampling happens \emph{after} the validation of blocks by the committee, i.e., past the 4-second deadline.
Completing sampling after committee validation may reverse validation decisions due to the delayed detection of blob data unavailability.
This requires modifications to Ethereum's consensus to account for such decision-revert possibilities.
This also impacts finality (i.e., how long before a block can be considered immutable) as pending data availability verifications delay decisions.
Finally, the possibility of reverting past consensus opens the door to new attacks based on \emph{ex-ante} reorganizations~\cite{d2024recent}.

\mypara{Contribution}
We demonstrate that DAS can be integrated with the existing Ethereum consensus while meeting the Danksharding objectives.
Dissemination and sampling of blob data can occur within the first four seconds of a consensus slot.
This allows the committee to confirm blob data availability and block data correctness simultaneously and removes the need to adapt Ethereum's consensus to delayed availability decisions.

We present \sysname, a peer-to-peer protocol that supports DAS in Ethereum.
\sysname builds upon the following key features:

\begin{compactitem} 

    \item 
    It aligns with recent Ethereum evolutions, including Proposer-Builder Separation (PBS)~\cite{ethereum2024roadmap_pbs,heimbach2023ethereum}, which introduces powerful builders responsible for preparing block and blob data and ordinary proposers elected by Proof-of-Stake consensus~\cite{merge-ethereum}.
    \sysname leverages builders for efficient \emph{seeding} of blob data. 
    
    \item
    It uses Ethereum nodes to host and sample blob data using peer-to-peer interactions.
    In contrast to other proposals~\cite{eip-7594,subnetdas,fulldas}, \sysname employs direct (one-hop) communication, using connectionless networking (UDP). 
    Interactions adapt to nodes' unavailability and faults, meeting the 4-second deadline in adverse environments or under inconsistent network views by different participants. 
    
    \item
    \sysname supports Ethereum's objectives of openness, decentralization, and scalability.
    Nodes' and builders' bandwidth requirements align with the typical capacities of home servers and cloud instances, and do not increase with system size.
    
\end{compactitem}

We implement \sysname over \texttt{libp2p}~\cite{libp2p}, the network stack of the Ethereum Geth client~\cite{geth}, and deploy 1,000 nodes on an 80-server cluster using representative emulated WAN latencies.
Additionally, we utilize a simulator whose results are cross-validated against prototype deployments.
This enables us to confidently explore results for up to 20,000 nodes.

Our evaluation shows that \sysname meets the 4-second sampling deadline at all nodes at moderate scales and for the vast majority of nodes at large scales, while maintaining low load on builders and nodes. In contrast, baseline solutions based on GossipSub~\cite{gossipsub} or the Kademlia DHT~\cite{maymounkov2002kademlia} do not scale as well, incurring higher overhead and failing to meet the 4-second deadline even at moderate network sizes. Experiments involving a significant fraction of unresponsive nodes and inconsistent views further demonstrate that \sysname's operations are robust against faults, meeting the 4-second deadline for the majority of nodes even when up to 50\% of the nodes are misbehaving, and systematically detect data unavailability. \er{TODO add more highlights of the evaluation here.}






\mypara{Outline}
This paper is organized as follows.
We present preliminaries about Ethereum, layer-2 protocols, and PBS (\Cref{sec:background}).
We detail the DAS principles and the \emph{Danksharding} roadmap and analyze the associated networking and communication requirements (\Cref{sec:das}).
We present our model and assumptions, and detail our design objectives (\Cref{sec:overview}).
\sysname uses a deterministic assignment of blob data to nodes (\Cref{sec:assignment}).
It operates in three phases, from the \emph{seeding} of blob data by a builder to nodes \emph{consolidation} of this data and its \emph{sampling} (\Cref{sec:pandas_phases}).
\sysname uses direct and efficient but unreliable UDP communications.
An adaptive fetching protocol arbitrates between request redundancy and time constraints (\Cref{sec:fetching}).
We evaluate \sysname and compare it to baselines (\Cref{sec:evaluation}).
We discuss our results (\Cref{sec:discussion}) before covering related work (\Cref{sec:related}) and concluding (\Cref{sec:conclusion}).

%% file: sections/background.tex
\vspacebeforesection
\section{Preliminaries}
\vspaceaftersection
\label{sec:background}


We provide an overview of Ethereum, its consensus, the Proposer-Builder Separation principle, and layer-2 protocols.

\mypara{Ethereum}
Ethereum is an open blockchain using Proof-of-Stake (PoS) consensus~\cite{merge-ethereum}.
Holders of ETH, Ethereum's virtual currency, can lock 32~ETH (their \emph{stake}) or more to operate a \emph{validator}, i.e., a virtual entity participating in the validation of new blocks.


Time in Ethereum is divided into slots of 12 seconds and epochs of 32 slots.
In every slot, a new block is added to the blockchain.
A subset of validators is deterministically selected to participate in each consensus slot.
One of them, the \emph{proposer}, is responsible for forming and spreading a new block. 
Some validators produce \emph{attestations} of this new block, while others collect these attestations and publish aggregate decisions.
As a result, consensus is split into three phases: 
  (1)~the broadcast of a new block and its verification by the committee;
  (2)~the propagation and collection of attestations;
  and
  (3)~the generation and broadcast of aggregate decisions.
Each phase accounts for a third of the slot duration, i.e., $12/3=4$ seconds.


Servers called full nodes, or simply ``nodes'' for the rest of this paper, participate in the Ethereum network.
Nodes can, but do not have to, host validators.
A node is identified by its IP address and a public key, which are shared through \emph{Ethereum Node Records} (ENR) propagated through the network and stored in the underlying Kademlia DHT~\cite{maymounkov2002kademlia,krol2024disc}.
While nodes can collect all ENRs by crawling the DHT~\cite{discv4-dns-lists,nebuladhtcrawler,cortes2021discovering,codexstoragedhtcrawler}, the association between a node and a specific validator should not be public~\cite{heimbach2025deanonymizing}.
Deanonymizing the link between the two leads to security threats such as DDoS or targeted abuse of slashing mechanisms~\cite{pavloff2024byzantine}.
All nodes, whether they host a validator or not, use the consensus committee's aggregate decisions to determine whether a block is accepted.


The dissemination of new blocks, attestations, and aggregate decisions is supported by Gossipsub~\cite{gossipsub}, a peer-to-peer overlay that enables multi-hop, controlled flooding of data. 

\begin{figure}[t]
    \centering
    \includegraphics[scale=\omnifiguresscalingfactor]{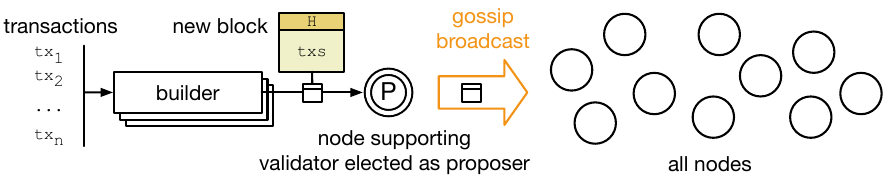}
    \vspacebeforecaption
    \caption{%
    Proposer-Builder Separation (PBS).
    The proposer, elected based on stake, selects a block among those prepared by builders.
    The block is broadcast by gossip to all nodes.
    }
    \vspaceaftercaption
    \label{fig:pbs_principle}
\end{figure}

\mypara{Proposer-Builder Separation}
Forming new blocks is increasingly computationally expensive, particularly with the rising importance of \emph{Maximal Extractable Value} or MEV~\cite{gramlich2024maximal}.
Any node hosting a validator that can be elected as a proposer would need to provision a powerful server.
Preventing low-stake participants unable to do so from participating in consensus leads to the concentration of power among a limited number of actors.
Proposer-Builder Separation (PBS), illustrated in \Cref{fig:pbs_principle}, addresses this risk by separating the role of \emph{building} a block and the role of \emph{proposing} it for consensus.
This enables a few dedicated builders to form new blocks while maintaining decentralized consensus among many lightweight nodes that host validators.
With PBS, the node supporting the proposer selects one of the blocks prepared by builders. 
The block is then broadcast to all nodes using Gossipsub.
Today, PBS is responsible for around 90\% of Ethereum block creation~\cite{mev-watch,heimbach2023ethereum}, principally through the MEV-Boost network~\cite{mev-boost}.
Builders receive block construction fees for blocks selected by proposers and accepted by consensus; therefore, they are incentivized to produce correct blocks.

\mypara{Layer-2 protocols}
The throughput of Ethereum's chain (i.e., layer-1) is limited to the number of transactions that can fit in a block. 
Layer-2 protocols move some of the transaction handling and validation processes to a separate layer while benefiting from the security and decentralization of the layer-1 chain.
There exist many variants of layer-2 protocols~\cite{gangwal2023survey,jourenko2019sok}.

Rollups are exemplary layer-2 solutions that process transactions off-chain.
They publish transaction state in a compressed form, together with a commitment to this state, via a call to a smart contract in a regular layer-1 transaction.
Rollup variants include optimistic ones~\cite{ethereum2024optimistic_rollups} posting compact hashes of transactions' states, e.g., Arbitrum~\cite{kalodner2018arbitrum} and Optimism/Bedrock~\cite{optimism}, and ZK rollups~\cite{ethereum2024zk_rollups} posting zero-know\-ledge proofs of validity, e.g., ZkSync~\cite{zksync} or Polygon~\cite{polygon}.
The volume of layer-2 transactions that can be \emph{anchored} to the layer-1 chain is directly linked to the supported volume of blob data.
This data needs to be available for a sufficient time for a protocol's participants to verify it (e.g., verifying the ZK proof~\cite{ethereum2024zk_rollups,zksync,polygon} or generating a fraud proof~\cite{ethereum2024optimistic_rollups,optimism,kalodner2018arbitrum}).
Unlike regular layer-1 transactions, layer-2 data does not need to persist indefinitely nor be verified for correctness by layer-1 nodes.

%% file: sections/das.tex
\vspacebeforesection
\section{Data Availability Sampling}
\vspaceaftersection
\label{sec:das}

\begin{figure}[t!]
    \centering
    \includegraphics[scale=\omnifiguresscalingfactor]{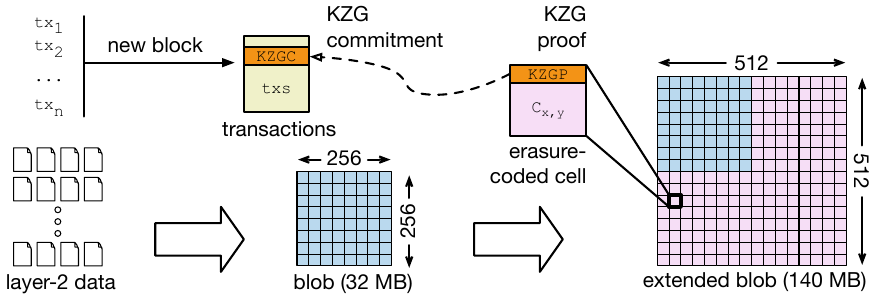}
    \vspacebeforecaption
    \caption{%
	Builder preparatory operations for DAS.
	32~MB of data is aggregated in a blob of $256\times 256$ cells, extended to $512\times 512$ cells using erasure coding.
	Each resulting cell includes a proof (KZGP) linking it with the Kate-Zaverucha-Goldberg commitment (KZGC) in the corresponding blob-carrying transaction.
  }
    \vspaceaftercaption
    \label{fig:das_principle}
\end{figure}

Ethereum's current mechanism for attaching layer-2 data to layer-1 blocks is EIP-4844 (\emph{Proto-Dank\-shar\-ding})~\cite{ethereum_protodanksharding_eip4844}.
It attaches a limited number of data \emph{blobs} (binary objects) to each block.
This blob data is broadcast to all nodes.
Nodes hosting committee members must validate the corresponding commitments contained in blob-carrying transactions.
To keep the costs of operating a node reasonable and preserve decentralization, EIP-4844 limits the number of 128-KB blobs to 3 (on average) to 6 (maximum) or 0.375 to 0.75~MB of data. 

\emph{Danksharding}~\cite{ethereum2024roadmap_danksharding} is a roadmap towards much more (32~MB) blob data attached to each block.
It is intimately linked to PBS and relies on builders' computational and networking power to collect, aggregate, and share layer-2 data.
With such volumes, fully disseminating blob data to all nodes is no longer realistic.
Instead, each node receives and stores a subset (shard) of it.
For an individual node, receiving a subset does not guarantee the availability of the \emph{complete} blob data.
Data Availability Sampling (DAS) enables this verification.
It consists of three phases.
First, the blob is extended using erasure coding. 
Second, each shard of extended blob data is distributed to a subset of nodes for hosting.
Third, nodes collect \emph{samples}, allowing them to consider the data they do not host available (or reconstructable) with an overwhelming probability.

\Cref{fig:das_principle} details the construction of blob data.
The blob aggregates 32~MB of data, split into cells of 512~B, organized as a $256 \times 256$ matrix.
Releasing only a subset of blob data is a \emph{data withholding attack}.
The base blob is highly amenable to such an attack, as sharing all but one cell makes some data unavailable, threatening the security of layer-2 protocols.
To prevent data withholding and allow data reconstruction after losses, the blob is \emph{extended} using a two-dimensional Reed-Solomon erasure code~\cite{wicker1999reed}.
Each row and column doubles in size but can now be reconstructed from any 50\% of its cells.
The resulting \emph{extended} blob is now a $512 \times 512$-cell matrix.
In addition to the 512~B of data, each cell includes a 48~B Kate-Zaverucha-Goldberg proof (KZGP)~\cite{kate2010constant}.
This proof links the cell's content to a commitment (KZGC) registered in a layer-1 blob-carrying transaction.
In total, the extended blob is $(512 \times 512) \times (512 + 48) = 140$ MB in size including 12~MB of KZGPs.

\begin{figure}[t!]
     \centering
     \includegraphics[scale=\omnifiguresscalingfactor]{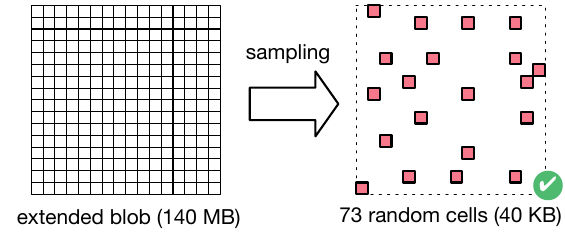}
     \vspacebeforecaption
     \caption{%
     Each node samples 73 randomly chosen cells. 
   }
     \vspaceaftercaption
     \label{fig:das_principle_sampling}
 \end{figure}

Following the dissemination of extended blob data, nodes verify its availability by attempting to download randomly chosen cells, as illustrated by \Cref{fig:das_principle_sampling}. 
Collecting more random samples means higher confidence in the availability or reconstructability of blob data.
The number of cells to sample depends on the maximum acceptable rate of false positives, i.e., of incorrectly determining availability, as we detail next.

\begin{figure}[t!]
    \centering
    \includegraphics[scale=\omnifiguresscalingfactor]{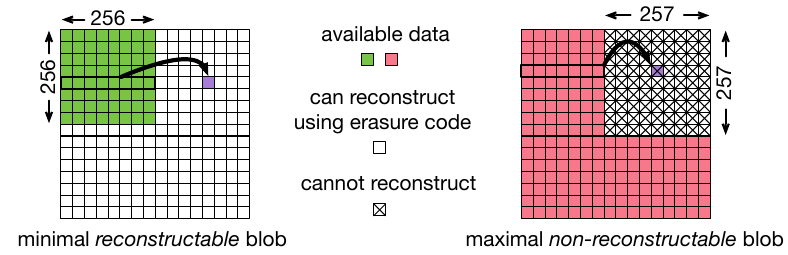}
    \vspacebeforecaption
    \caption{%
    The minimal data enabling reconstruction (left), and the maximal data preventing it (right).
  }
    \vspaceaftercaption
    \label{fig:das_reconstructable_vs_non_reconstrucable}
\end{figure}

The minimal amount of data necessary to enable reconstruction is half of the cells for either 256 distinct rows or 256 distinct columns, as illustrated by \Cref{fig:das_reconstructable_vs_non_reconstrucable}-left (note that collecting \emph{any} $256\times 256=65,536$ cells \emph{may not} provide this guarantee).
The maximal amount of data that can be shared while preventing reconstruction is the $512 \times 512$ matrix minus a $257 \times 257$ square sub-matrix, as illustrated by \Cref{fig:das_reconstructable_vs_non_reconstrucable}-right.
If a fraction $p$ of cells was not shared in the network, the probability of not hitting an unavailable cell with $s$ samples is $(1-p)^{s}$.
The false positive probability for availability sampling is, therefore, upper-bounded by $\prod_{i=0}^{s-1}{1-\frac{257\times257}{512\times 512-i}}$.
Discussions in the Ethereum community~\cite{collab_computations_das} suggest using $s=73$ samples, which gives an upper bound false positive probability lower than $10^{-9}$.
We use this value of $s=73$ in the remainder of the paper, corresponding to $73 \times 560 \text{B} = 40$~KB worth of samples collected per node.

%% file: sections/overview.tex
\vspacebeforesection
\section{\sysname: Objectives and Overview}
\vspaceaftersection
\label{sec:overview}

\sysname is a peer-to-peer protocol integrating DAS without requiring modifications to Ethereum.
It ensures that extended blob data is propagated to the network and sampled within 4 seconds of the block's creation.
This section presents our assumptions, details our objectives, and gives an overview of \sysname.

\vspacebeforesubsection
\subsection{Model and Assumptions}
\vspaceaftersubsection

This work is based on the following models and assumptions. 


\mypara{System model}
The network comprises $N$ nodes. 
Aligned with Ethereum, the system is open, 
but each node $n_i \in N$ is identified by an ID $i$, i.e., a cryptographic hash of its public key.
Nodes periodically advertise to store (and refresh) their ENR records in the underlying Kademlia DHT.
The ENR of a node contains its ID, public key, and contact information (IP and port).
Nodes can be reached directly using this contact information.


The assignment of validators to nodes must remain unknown~\cite{heimbach2025deanonymizing}.
Therefore, it must be impossible to distinguish nodes that host validators from those that do not.
To maintain decentralization, nodes must have commodity hardware and network requirements (i.e., a small home server with a 25~Mbps network connection~\cite{hardware_proposers}).

Dedicated builders propose blocks.
For every slot, the elected proposer selects one block from a builder $b$.
Builders have significantly better capacity and connectivity than nodes (e.g., a medium-range cloud instance with a recent multicore CPU and 10~Gbps network upload capacity).
The selected builder $b$ is responsible for sending extended blob data to the network of nodes. 
After that, nodes interact peer-to-peer to exchange this data for retrieval and DAS.



\mypara{Network views}
Each node, including builders, maintains a list of all nodes in the system as its \emph{view} $V$, i.e., $V_{b}$ is builder $b$'s knowledge of existing nodes, and $V_{n_1}$ is that of a node $n_1$.
Views are filled by periodically crawling the DHT~\cite{discv4-dns-lists}, which typically takes about a minute~\cite{nebuladhtcrawler,cortes2021discovering,codexstoragedhtcrawler}.
Views can be inconsistent (for any two nodes $n_1$ and $n_2$, we do not assume that $V_{n_1} = V_{n_2}$, and similarly for builders). 
They may also be incomplete ($V \cap N \subseteq N$) and contain departed nodes ($V - N \neq \emptyset$).
However, 
thanks to the periodic crawls, views constantly converge towards the actual set of nodes.

\mypara{Fault/attack model}
Nodes may crash (fail-silent~\cite{powell1995failure}) or refuse to answer incoming requests. 
Builders are rational and follow their economic interests.
They aim to obtain block construction rewards while spending as few resources as possible. 
A selected builder can attempt a \emph{data withholding} attack, i.e., avoid sharing some or all of the blob data to save on operational cost or because it did not produce it.
However, it does not attempt to send \emph{incorrect} data to the network, as doing so would be against its economic interests (i.e, this will be detected when checking KZGP and lead to no rewards, while still incurring bandwidth costs).



\vspacebeforesubsection
\subsection{Objectives}
\vspaceaftersubsection


The primary objective of \sysname is to ensure that dissemination and sampling happen within four seconds of creating a block, and layer-2 clients easily retrieve blob data.
We set the following goals:
\begin{itemize}
    \item \textbf{[Robustness]} Sampling must meet the 4-second deadline even with a large fraction of unresponsive nodes and/or when nodes and builders have inconsistent views.
    \item \textbf{[Scalability]} Timing guarantees must hold with increasing system size, and the load imposed on nodes and builders must remain compatible with hardware profiles recommended for decentralization~\cite{hardware_proposers}.
    \item \textbf{[Flexibility]} Participating entities may be free to implement local strategies for interacting with other system members, aligning with their financial incentives.
\end{itemize}

We target the \emph{tight fork-choice} rule~\cite{das_fork_choice_may24}, i.e., DAS sampling is required before attesting to a block by committee members: a block with valid transactions but unavailable data is attested as invalid.
As a result, we do not modify the consensus protocol beyond adding sampling as a verification step for nodes hosting committee member validators.
This contrasts with the \emph{trailing fork-choice} rule that postpones sampling to later, and requires non-trivial changes to consensus to be able to revert blocks with unavailable blob data.
Similarly, we do not wish to modify Ethereum's discovery protocols (i.e., the DHT holding ENRs) and assume nodes use unmodified crawl mechanisms to collect their views~\cite{nebuladhtcrawler,cortes2021discovering,codexstoragedhtcrawler}.


\vspacebeforesubsection
\subsection{\sysname in a nutshell}
\vspaceaftersubsection

\begin{figure}[t!]
    \centering
    \includegraphics[scale=\omnifiguresscalingfactor]{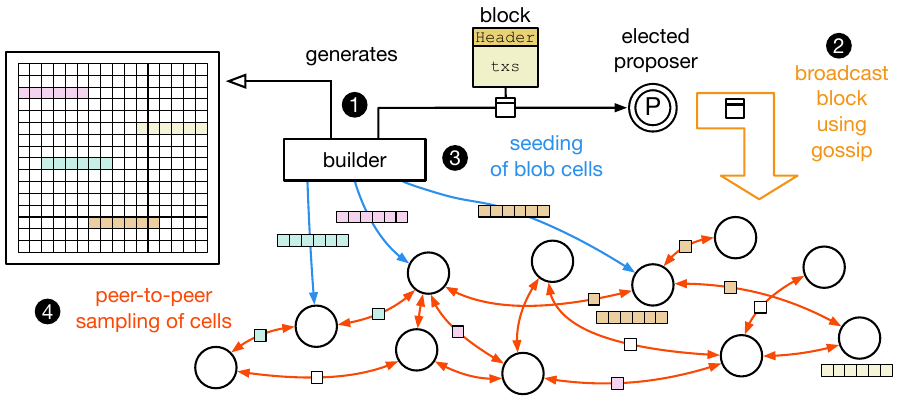}
    \vspacebeforecaption
    \caption{%
    Distributed interactions following the selection of a new block by the proposer (\ding{202}).
    In parallel to the gossip block dissemination (\ding{203}), the builder distributes extended blob data to nodes in the network (\ding{204}).
    All nodes interact peer-to-peer to \emph{consolidate} their assignment and collect random samples (\ding{205}).}
    \vspaceaftercaption
    \label{fig:overview_networking}
\end{figure}

The high-level principles of \sysname are illustrated by \Cref{fig:overview_networking}.
At the beginning of a slot, the node hosting the elected proposer selects a block from one of the builders (\ding{202}).
This block is disseminated via a dedicated, system-wide GossipSub channel (\ding{203}).
At the same time, the same node requests the builder to publish blob data in the network.
The builder \emph{seeds} the network with extended blob cells, using direct communication to nodes in its view (\ding{204}).
Every node is assigned a subset of cells that it must keep in \emph{custody} for the rest of the network.
The builder may send only a subset of this assigned data to each node directly.
To serve all assigned data, nodes fetch missing cells from other nodes through \emph{consolidation} (\ding{205}).
In parallel, nodes select 73 cells randomly and send requests to nodes whose responsibility includes them, implementing the \emph{sampling} phase.

\begin{figure}[t!]
    \centering
    \includegraphics[scale=\omnifiguresscalingfactor]{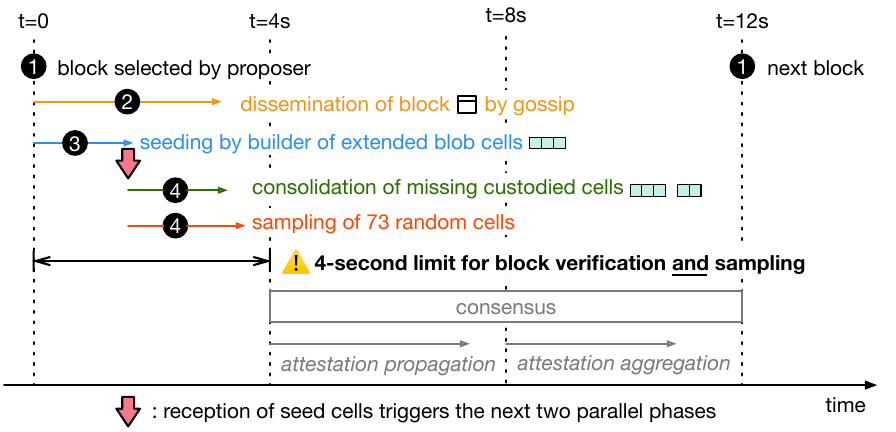}
    \vspacebeforecaption
    \caption{%
    Timeline of events within a slot.
    Starting from the selection of a new block by the proposer (time 0, \ding{202}), two concurrent processes start for nodes that must both terminate within 4 seconds: block dissemination (\ding{203}) and extended blob data dissemination (\ding{204}), consolidation, and sampling (\ding{205}). 
    }
    \vspaceaftercaption
    \label{fig:das_timeline}
\end{figure}

\Cref{fig:das_timeline} represents the timeline of operations.
The dissemination and verification of block and blob data are concurrent.
Nodes initiate consolidation and sampling when they receive their seed cells from the builder.
A node supporting an active validator can vote for a block if the block verification \emph{and} the data availability sampling are successful before the 4-second deadline.

As ENRs do not allow distinguishing between nodes supporting validators and nodes that do not, all correct nodes are expected to custody data as long as they are registered in the DHT.
We also assume all correct nodes perform DAS.
In particular, we avoid having only committee members performing DAS, as it would reveal the association between validators in the committee and nodes~\cite{heimbach2025deanonymizing}.

Communication between all actors in \sysname is based on one-way UDP networking with no signalling overhead (i.e., there is no establishment of connections or keep-alive messages).
We stress that all Ethereum nodes already use UDP in the discovery protocol required to join the network~\cite{discv5udp}. 
Blob data is public and, therefore, sent unencrypted, avoiding a time-consuming encrypted channel establishment.
Messages are authenticated with a digital signature using the recipient's public key.
KZGPs further allow the authenticity of the received blob data to be verified.
Peer-to-peer requests may fail silently due to packet loss or nodes that are unresponsive or have failed.
To alleviate this and meet the deadline, \sysname relies on builders adopting efficient seeding strategies, reconstructing cells using the erasure code, and nodes employing an adaptive fetching strategy that adapts request redundancy and aggressiveness to the available time budget.

Similarly, the impact of incorrect nodes that do not participate in custody and consolidation is mitigated by redundancy.
Incorrect nodes that forfeit the sampling phase only reduce the system load.

In the following sections, we detail the components of \sysname.
We start with the deterministic association between blob data and nodes (\Cref{sec:assignment}).
Then, we present the three phases of \emph{seeding}, \emph{consolidation}, and \emph{sampling} (\Cref{sec:pandas_phases}).
We finally detail the adaptive fetching strategy (\Cref{sec:fetching}).

%% file: sections/assignment.tex
\vspacebeforesection
\section{Cell to Nodes Assignment}
\vspaceaftersection
\label{sec:assignment}

The first component of \sysname is an assignment between blob data and nodes.
A function $\sigma(n_i)$ returns a list of cells from the 512$\times$512. 
Node $n_i$ is tasked with their custody, i.e., hosting and serving these cells for sampling queries and access by layer-2 participants.

\begin{figure}[t]
    \centering
    \includegraphics[scale=\omnifiguresscalingfactor]{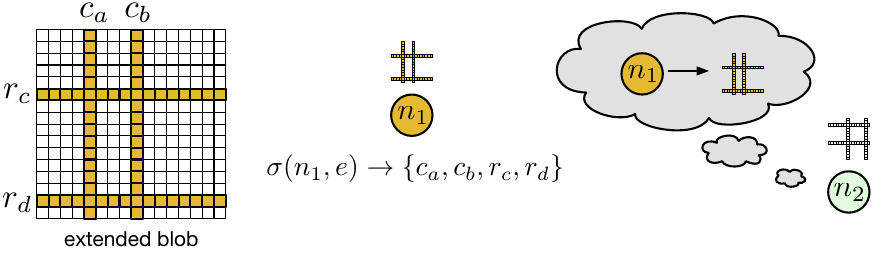}
    \vspacebeforecaption
    \caption{%
    Assignment $\sigma$ of cells to nodes.
    Node $n_1$ is assigned columns $c_a$ and $c_b$ and rows $r_c$ and $r_d$ in epoch $e$.
    Any node $n_2$ that knows $n_1$ deterministically determines its assigned cells, regardless of the rest of its view $V_{n_2} \setminus n_1$.
    }
    \vspaceaftercaption
    \label{fig:assignment}
\end{figure}

All nodes and builders know $\sigma$, as illustrated by \Cref{fig:assignment}.
We set two requirements for $\sigma$: it must be \emph{deterministic} and \emph{short-lived}.
Determinism means $\sigma(n_i)$ must be computed identically by two nodes $n_a$ and $n_b$ even if $V_{n_a} \neq V_{n_b}$.\footnote{Using consistent hashing, as in DHTs, does not meet this requirement: if $n_a$ knows a node $n_c$ that $n_b$ does not know, $n_a$ may associate cells to $n_c$ that $n_b$ associates to $n_a$.}
Short-liveness means that the assignment must change periodically and be unpredictable.
This prevents the emergence of attacks based on eclipsing nodes in charge of specific cells~\cite{wust2016ethereum,marcus2018low} or censorship of specific data~\cite{sridhar2024content}.


Even though adjacent cells likely contain data for distinct layer-2 protocols, storing them together on the same node favors efficient reconstruction, as it requires fetching multiple cells from the same row or column.
Thus, \sysname assigns complete rows and columns to each node.
The number of rows and columns assigned to each node is a globally known parameter.
By default, we use eight distinct rows and columns per node.
Each node hosts $8\times 512 + 8\times (512-2)$ cells, i.e., $8,176 \times (512 + 48) = \simeq 4.4$ MB of data.\footnote{While this is the expected amount of data stored by a node to enable DAS, nothing prevents collecting more. Typically, nodes participating in layer-2 protocols may obtain all relevant data and cache it for other participants in their network.}


To enable determinism and short-liveness, the assignment $\sigma$ is a pseudo-random sortition. 
This is the exact mechanism used to select committees in the Ethereum consensus.
For every epoch, a globally verifiable, pseudo-random sortition decides which nodes will be members of committees or proposers in each slot.
This decision uses a pseudo-random number generator (PRNG) and an \emph{epoch seed} known one epoch in advance (32 slots, $\simeq$6.4 minutes) from a combination of random values proposed by validators (i.e., the ``RANDAO'' state~\cite{randao}). 
\sysname builds upon this existing mechanism by seeding the assignment function $\sigma$ for an epoch $e$ with its corresponding epoch seed ${\mathit{s}_e}$.
We extend the definition of $\sigma$ to include the epoch number, i.e., function $\sigma(n_i,e)$ generates eight distinct rows and eight distinct columns for $n_i$ using a PRNG seeded by ${\mathit{s}_e}$.

%% file: sections/pandas_phases.tex
\vspacebeforesection
\section{\sysname Protocol Phases}
\vspaceaftersection
\label{sec:pandas_phases}

We detail the \sysname protocol phases: seeding, consolidation, and sampling, illustrated in Figures~\ref{fig:overview_networking} \& \ref{fig:das_timeline}.
The latter two are concurrent.


\vspacebeforesubsection
\subsection{Seeding phase}
\vspaceaftersubsection
\label{sec:pandas_phases:seeding}

The interactions start with an initial \emph{seeding} phase.
This phase starts when a proposer selects a block from a builder $b$.
In parallel to sending the block to the network via gossip, the proposer asks $b$ to seed the corresponding blob data to the network.

All nodes know the proposer's identity and public key before the slot starts.
However, they do not know who $b$ is.
Due to the strict time constraints, nodes cannot wait to receive the block via gossip to learn this information and start accepting blob data.
To allow nodes to distinguish legitimate blob data, the proposer provides the builder with a digital signature binding $b$'s identity (including its IP address) to the proposer's private key.
This signature is attached to every seeding message.\footnote{While the proposer's signature allows verifying the legitimacy of $b$, the correctness of the received cells' KZGP from $b$ cannot be checked against the KZGC before receiving the block and its blob-carrying transactions. It is, however, not in the builder's interest to send fake blob data that will eventually cause it to lose the rewards.} 

An objective of \sysname is flexibility, i.e., the possibility for different actors to implement various strategies. 
This principle applies to blob seeding strategies.
Builders are rational; their interest is in operational costs, particularly outgoing bandwidth.
They also wish to maximize profits through block production rewards, which depend on the success of DAS.


A naive approach could be to have the selected builder $b$ send all cells in $\sigma(n,e)$ to every node $n \in V_b$.
The necessary outgoing bandwidth now depends on the size of the builders' view, close to or equal to that of the entire network.
With $\simeq$4.4~MB per node (eight rows and eight columns) and, say, 10,000 known nodes, the necessary bandwidth budget is 42.9~GB (343.7~Gb). 
With a 10~Gbps connection, as available with modern, medium-end cloud instances, the process takes more than 30 seconds, largely missing the 4-second deadline.

A better approach is to send a fixed amount of data to the network and determine a level of redundancy for the cells within each row and column.
For a row (or column) $x$, $b$ decides which cells of $x_{1}, \dots, x_{512}$ to send to the network and with what degree of redundancy.
It dispatches these cells to the nodes assigned to $x$ in the current epoch $e$ that it knows, i.e., $V_b(x) = \{ n \in V_b \where x \in \sigma(n,e) \}$.
Each node in $V_b(x)$ receives only a \emph{subset} of its assigned cells.
Therefore, each node must still fetch the missing cells from its peers, a process we call \emph{consolidation} that we detail in the following subsection.

\begin{figure}[t]
    \centering
    \includegraphics[scale=\omnifiguresscalingfactor]{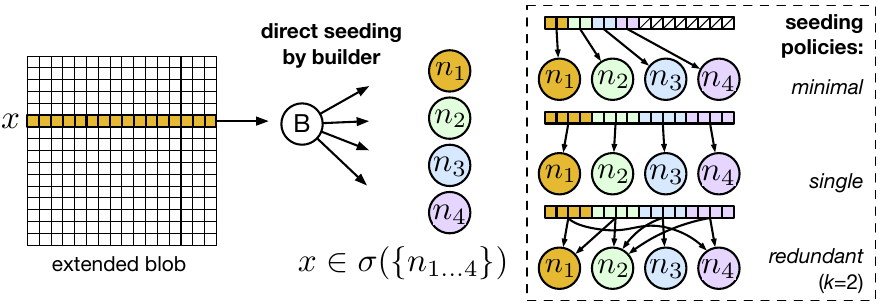}
    \vspacebeforecaption
    \caption{%
    Three seeding policies.
    The \emph{``minimal''} policy splits the first half of each row or column amongst known peers having it in their assignment.
    The \emph{``single''} policy splits the entire row or column.
    The \emph{``redundant''} policy shares each split to $k$ nodes. 
    }
    \vspaceaftercaption
    \label{fig:seeding_policies}
\end{figure}

\mypara{Seeding policies}
\Cref{fig:seeding_policies} illustrates three example policies.

A \emph{``minimal''} policy sends a single copy of \emph{half} of the cells of $x$, i.e., $x_{1}, \dots, x_{256}$ (i.e., the minimal amount of data necessary to reconstruct the row or column).
The builder splits $x_{1}, \dots, x_{256}$ into $|V_b(x)|$ parcels of adjacent cells and distributes them randomly to up to 256 nodes in $V_b(x)$.
It repeats the process for all rows and columns.
The total amount of data sent out is $256 \times 256 \times (512 + 48) = 35$~MB of data.
This strategy is exceptionally fragile to message loss.
We primarily use it as a baseline for the builders' costs. 

A second, \emph{``single''} policy leverages the redundancy allowed by the erasure code.
It operates similarly to the minimal policy but sends a single copy of \emph{all} of the cells of each row or column $x$, split to up to 512 nodes in $V_b(x)$.
In total, it sends out the size of the extended blob, i.e., 140~MB.
This strategy's rationale is that even if half of the cells are lost, nodes can still reconstruct the row or column using the erasure code.

The third, \emph{``redundant''} strategy, adds further redundancy by sending $k$ copies of each cell.
It starts from the single policy, splitting the cells of $x$ between nodes in $|V_b(r)|$.
Then, each parcel is further assigned to $k-1$ randomly selected distinct nodes in $|V_b(r)|$.
We use $k=8$ by default for this strategy.
The outgoing bandwidth usage for the builder is, therefore, 1,120~MB = 1.09~GB of data.
This is precisely the volume of data a builder would use to send every row and every column to GossipSub~\cite{gossipsub} channels for dissemination, as the typical fanout for the root of GossipSub dissemination trees is eight peers~\cite{gossipsub_spec}.
As using GossipSub channels is the approach proposed by concurrent designs to integrate DAS in Ethereum~\cite{eip-7594,subnetdas,fulldas}, this allows us to compare the performance of \sysname under the same resource utilization.


\vspacebeforesubsection
\subsection{Consolidation phase}
\vspaceaftersubsection
\label{sec:pandas_phases:consolidation}

\begin{figure}[t]
    \centering
    \includegraphics[scale=\omnifiguresscalingfactor]{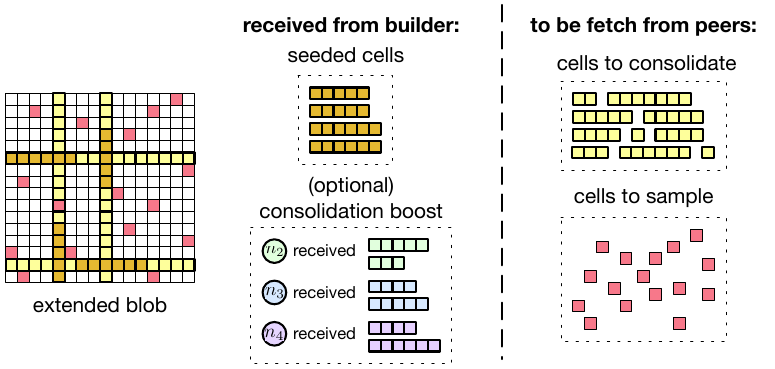}
    \vspacebeforecaption
    \caption{%
    Consolidation and sampling phases must fetch cells from other nodes.
    Information from the builder consists of initial cells and an optional consolidation boost map informing $n$ what other nodes received as seed data.
    }
    \vspaceaftercaption
    \label{fig:consolidation_sampling}
\end{figure}

The objective of the second phase, consolidation, is for a node $n$ to rapidly get hold of all the cells of rows and columns assigned by $\sigma(n,e)$.
Thanks to the erasure code, collecting half of the cells of a given row or column is sufficient to consolidate it.

Consolidation starts at $n$ upon reception of seed cells from the builder $b$.
If $n$ receives a request from another node linked to a slot for which it has not yet received its seed cells, it activates a timer (we use a default value of 400~ms).
Consolidation starts without seed data if the timer expires before $n$ receives cells from $b$ because of a packet loss or because $b$ does not know $n$ yet (i.e., $n \not\in V_b$). 

The orchestration and timing of requests for fetching missing cells are delegated to \sysname's fetching strategy, shared with the sampling phase, and detailed in the next section.

To fetch cells missing for consolidation, $n$ contacts peers with overlapping rows and columns, $C_n(x) = \{ n' \in V_n \where x \in \sigma(n,e) \land x \in \sigma(n',e) \}$. 
In a view with 10,000 nodes assigned eight rows and eight columns each, each row or column is assigned to $\frac{10000\times (8+8)}{512\times2}\simeq156$ nodes on average.
Thus, $C_n(x)$ contains about 624 peers.
Depending on the builder's seeding strategy, each may have received only a subset of the data.
Asking many peers for cells may increase the chances of ``hitting'' the ones that received the needed cells via seeding, but it leads to many duplicates.
In contrast, asking only a few random peers may require the selected nodes to finish their consolidation to respond, leading to a lengthy response delay. 

A fast and effective consolidation aligns with the economic interests of the builder.
If consolidation is fast and efficient, it improves the odds that sampling finishes on time, and the builder may afford to send less data to the network.
We improve these two factors with \emph{consolidation boosting}, as illustrated by \Cref{fig:consolidation_sampling}.
The builder $b$ attaches to the seeding message to $n$ a map $\mathit{CB}$.
For every row and column $x \in \sigma(n,e)$, $\mathit{CB}(x)$ lists the cells received by other nodes $n' \in V_b$ where $x \in \sigma(n',e)$.
The consolidation boosting map $\mathit{CB}$ allows $n$ to know which nodes are likely to receive specific cells faster and prioritize them for the requests.

\vspacebeforesubsection
\subsection{Sampling phase}
\vspaceaftersubsection
\label{sec:pandas_phases:sampling}

The third phase of \sysname is the \emph{sampling} phase.
It starts at the same time as consolidation and takes place concurrently.

Node $n$ randomly selects 73 cells to sample.
This selection must be unpredictable (i.e., unlike $\sigma$).
For every sample, $n$ can determine the nodes hosting an intersecting row or column in $V_n$.
In a 10,000-node network, 156 nodes on average can have a copy of a given cell.
The selection of targets for sampling and the orchestration and scheduling of requests are delegated to \sysname's fetching algorithm that we describe next.

%


%% file: sections/fetching.tex
\vspacebeforesection
\section{Adaptive Fetching}
\vspaceaftersection
\label{sec:fetching}

Both consolidation and sampling require fetching cells from other nodes.
This collection is handled as a single task by the adaptive fetching algorithm we detail in this section.

The fetching algorithm inputs a set of cell identifiers, as illustrated by \Cref{fig:consolidation_sampling}.
In addition, it may receive a consolidation boost map $\mathit{CB}$.
The algorithm aims to retrieve all cells before the 4~s deadline.

Target nodes are identified by a node $n$ from its view $V_n$ using the assignment $\sigma$.
Some of these nodes may be offline or unresponsive.
As \sysname uses connectionless communications using UDP, the network may silently lose queries or response messages.
Sending queries for cells sequentially bears the risk of missing the deadline.
On the contrary, sending queries to multiple nodes that hold a copy of each desired cell generates a swarm of messages in the network.
This leads to congestion risks, suggesting the need for a compromise between cautious fetching initially and a more aggressive approach with more redundant queries as the deadline approaches.

Consolidation processes at different nodes are executed concurrently.
A queried node $n_i$ may have a cell $c$ in its assignment $\sigma(n_i,e)$ but have not yet received it from the builder or through consolidation.
Nodes receiving a query for assigned cells they do not yet have buffer this query and respond when they can (if the cells are never received, they never respond, i.e., there is no negative acknowledgment).
Thus, a querying node may allow sufficient slack time for queried nodes to respond, particularly early in the slot. 

\begin{figure}[t]
    \centering
    \includegraphics[scale=\omnifiguresscalingfactor]{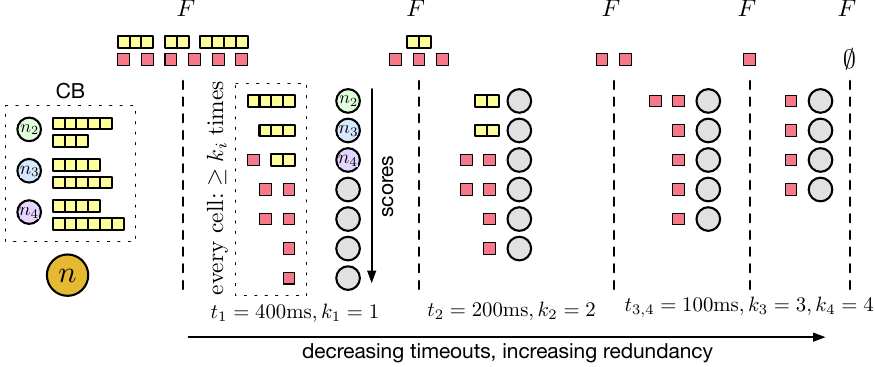}
    \vspacebeforecaption
    \caption{%
    Node $n$ determines a set of nodes to query at each round.
    The adaptive strategy adjusts the redundancy of queries (i.e., the number of nodes queried for each missing cell) and the timeout before the next round.
    }
    \vspaceaftercaption
    \label{fig:fetching}
\end{figure}

\begin{algorithm}[t]%
    \caption{Adaptive fetching at node $n$ in epoch $e$.}
    \label{alg:fetching}

    \begin{algorithmic}[1]%
     \scriptsize
        \Procedure{Fetch}{$F$, $\mathit{CB}$} \label{alg:fetching:fetch}
        
            \State $Q \gets V_{n}$; $i \gets 1$ \Comment{\textbf{Q}ueryable nodes and round number} \label{alg:fetching:Q}
            
            \While{$F \neq \emptyset \wedge i < i_{\text{max}}$}  \label{alg:fetching:mainWhile} \Comment{Until all fetched or too many rounds}
                
                \For{\textbf{each} $q \in Q$} \label{alg:fetching:scoringStart} \Comment{Assign scores to queryable nodes}
                    \State $q_\text{cells} = \{ \sigma(q,e) \subseteq F \}$ \Comment{Cells of interest ...} \label{alg:fetching:scoring:base_start}
                    \State $q_\text{score} = \| q_\text{cells} \|$ \Comment{... score is number of cells} \label{alg:fetching:scoring:base_stop}
                    
                    \If{$\mathit{CB}_q \neq \emptyset$} \Comment{If node in consolidation boost map ...} \label{alg:fetching:scoring:consolidation_start}
                        \State $q_\text{score} \gets q_\text{score} + (| F \cap \mathit{CB}_q |) \times \mathit{cb\_boost} $ \\ \Comment{... boost score for each cell received by seeding} \label{alg:fetching:scoring:consolidation_stop}
                    \EndIf
                \EndFor
                
                \State \textbf{sort} $Q$ by decreasing node score as $q_1,\dots,q_{|Q|}$ \label{alg:fetching:scoringStop}
                
                \State $P = \emptyset$; $U = F$; $j=1$ \Comment{Query \textbf{P}lan and cells \textbf{U}nder redundancy} \label{alg:fetching:planningStart} 
                
                \While{$U \neq \emptyset \wedge j \leq |Q|$}  \label{alg:fetching:whileStart} \Comment{Should/can plan more queries}
                    \If{$(q_j.\text{cells} \cap U) \neq \emptyset$} \Comment{At least one cell of interest} \label{alg:fetching:havemissingredundancy} 

                        \State $P \gets P \cup (q_j,q_j.\text{cells} \cap U)$ \Comment{Plan query} \label{alg:fetching:planquery} 
                        \State $U \gets \{ c \in F \where | \{ p \in P \where c \in p.\text{cells} \} | < k_i \}$ \label{alg:fetching:updateU} 
                        \\ \Comment{Update set of cells with insufficient redundancy}
                    \EndIf               
                    
                    \State $j \gets j+1$ \Comment{Consider next node in sorted $Q$} \label{alg:fetching:planningStop} 
                \EndWhile
                
                \For{\textbf{each} $p$ in $P$} \Comment{Send out queries from the query plan} \label{alg:fetching:executionStart} 
                    \State \Call{queryCells}{$p.\text{node}$,$p.\text{cells}$} \Comment{UDP async. query}
                    \State $Q \gets Q \setminus p.\text{node}$ \Comment{Nodes are queried only once} \label{alg:fetching:executionStop} 
                \EndFor
                
                \State $\Call{Sleep}{\algvar{t_i}}$; $i \gets i+1$ \Comment{Wait before next round} \label{alg:fetching:waitTimeout}
            \EndWhile
            
            \State \textbf{return} $(F \neq \emptyset)$ \Comment{Success if all cells fetched within round limit} 
            
        \EndProcedure

        \Procedure{UponReceive}{$C$} \Comment{Receiving a set of cells $C$} \label{alg:fetching:receiveStart} 
            \State $F \gets F \setminus C$ \Comment{Receive new cells}
            
            \While{$\exists$ row or column $x$ with $[256:512)$ cells} \Comment{Can use code} \label{alg:fetching:receive:check_reconstruction} 
                \State $x \gets$ \Call{Reconstruct}{$x$,$R$} \Comment{Reconstruct full row/column} \label{alg:fetching:receive:perform_reconstruction}
                \State $F \gets F \setminus \{ c \in x \}$ \Comment{No need to fetch reconstructed cells} \label{alg:fetching:receiveStop} 
            \EndWhile
            
        \EndProcedure

    \end{algorithmic}%
\end{algorithm}

\mypara{Fetching algorithm}
Fetching operates in rounds, as illustrated in \Cref{fig:fetching}.
It adapts query redundancy and timeouts as time progresses and the deadline nears.
For this purpose, each round $i$ is associated with a timeout $t_i$ and a redundancy factor $k_i$. 
%
\Cref{alg:fetching} details the process at a node $n$.
The \textsc{Fetch} procedure receives a set of cells to fetch $F$ and an optional consolidation boost map $\mathit{CB}$ (\Cref{alg:fetching:fetch}).
Node $n$ considers as \emph{queryable} nodes all of its view $V_n$ upon the initial call to \textsc{Fetch}, saved as a working copy $Q$ (\Cref{alg:fetching:Q}).
Any node in $Q$ will be queried at most once.
The fetching process is in three steps: \emph{scoring}, \emph{planning}, and \emph{execution}.

In the scoring step (\Crefrange{alg:fetching:scoringStart}{alg:fetching:scoringStop}), queryable nodes in $Q$ are assigned a score, i.e., the number of their assigned cells still missing for $n$ (\Cref{alg:fetching:scoring:base_start,alg:fetching:scoring:base_stop}).
If a consolidation boost was received, nodes are given a score boost of $\mathit{cb\_boost}$ for each cell declared as seeded by the builder and missing from $F$ (\Cref{alg:fetching:scoring:consolidation_start,alg:fetching:scoring:consolidation_stop}). 
The set of queryable nodes is then sorted by decreasing score values (\Cref{alg:fetching:scoringStop}).

The planning step (\Crefrange{alg:fetching:planningStart}{alg:fetching:planningStop}) prepares the set of queries as a set $P$.
Each planned query $p \in P$ is associated with a node $p.\text{node}$ and queried cells $p.\text{cells}$.
Each missing cell from $F$ must be queried from $k_i$ nodes.
Starting from the node with the highest score, $q_1$, the step greedily selects nodes with cells of interest as long as this criterion is not met.
For this, it maintains a set $U$ listing the cells for which insufficient redundancy currently exists in $P$.
A node $q_j$ is planned to be queried for cells with insufficient redundancy (\Cref{alg:fetching:planquery}), before updating $U$ (\Cref{alg:fetching:updateU}).

Finally, the execution step sends out the queries asynchronously (\Crefrange{alg:fetching:executionStart}{alg:fetching:executionStop}) before waiting for $t_i$~ms before the next round.
A queried node is removed from $Q$ and is not used again.
Upon correct reception by the target node, the handler either responds with the queried cells if all are available or buffers the query for a delayed reply.
The response is received by the \textsc{UponReceive} function as a set of cells $C$ (\Crefrange{alg:fetching:receiveStart}{alg:fetching:receiveStop}).\footnote{For clarity, we omit in \Cref{alg:fetching} the verification checks performed when receiving $C$ (e.g., verifying the cells KZMPs if/when the block header is available).}
When receiving new cells, the algorithm checks if an incomplete row or column now contains 256 or more cells (\Cref{alg:fetching:receive:check_reconstruction}) and, if so, reconstructs them (\Cref{alg:fetching:receive:perform_reconstruction}).
\rs{The notation might be a bit confusing. $F$ is a parameter of the Fetch function but its value is modified by the UponReceive function during execution} \er{TODO if space allows: add a new variable $M$ for Missing, copy it from $F$ upon a call to \algvar{Fetch}, and modify $M$ in the \algvar{UponReceive} method.}

\mypara{Default parameters}
The fetching algorithm is primarily parameterized by the round durations and query redundancy vectors $t$ and $k$, as well as the score boost for consolidation $\mathit{cb\_boost}$.
We use the following universal parameters, but stress that nodes could select them differently, e.g., based on local connectivity.

In the first round, $i=1$, the strategy aims to maximize the number of cells received (and reconstructed) using as few messages as possible (i.e., $k_1=1$).
We use a duration of $t_1=400 \mathit{ms}$ based on estimated time for the builder to send out initial cells and on inter-node latencies, as we will detail in \Cref{sec:evaluation}.
In subsequent rounds, we reduce this time by half but no lower than $100 \mathit{ms}$, i.e., $t_2 =200 \mathit{ms}$ and $\forall j \geq 3, t_j =100 \mathit{ms}$ (up to $t_{50}$).
Similarly, we increase the aggressiveness of queries by increasing the redundancy factor by two every round until a maximum of 10, i.e., $r_2=2, r_3=4, \dots, \forall j \geq 6, r_j=10$.
Finally, we set $\mathit{cb\_boost}=10,000$ to give an overwhelming advantage to nodes with seeded cells of interest.
\rs{Is $i_{max}$ set such that the four seconds are fully used?} \er{I added $t_{50}$ as the max but Matthieu can tell what is used in the implementation.}

%% file: sections/evaluation.tex
\vspacebeforesection
\section{Evaluation}
\vspaceaftersection
\label{sec:evaluation}

We structure our evaluation around the following claims:
\begin{itemize}
    \item \textbf{C1:} \sysname completes DAS within 4~s and supports the tight-fork choice rule under Danksharding requirements.
    \item \textbf{C2:} \sysname bandwidth requirements for nodes are below Ethereum suggestions for decentralization (25~Mbps~\cite{hardware_proposers}) and, for builders, below typical cloud offerings (10~Gbps).
    \item \textbf{C3:} \sysname satisfies \textbf{C1} even under a high percentage of unresponsive nodes and with highly inconsistent views.
    \item \textbf{C4:} \sysname satisfies \textbf{C1}--\textbf{C3} scaling up to 20,000 nodes. 
    \item \textbf{C5:} Relying on existing peer-to-peer overlays (Gossipsub~\cite{gossipsub} and Kademlia~\cite{maymounkov2002kademlia}) for DAS does not allow satisfying \textbf{C1}.   
\end{itemize}

 \begin{figure*}[t!]
    \hspace{-3em}
    \begin{subfigure}{0.18\textwidth} \centering
    \includegraphics[scale=\plotscale]{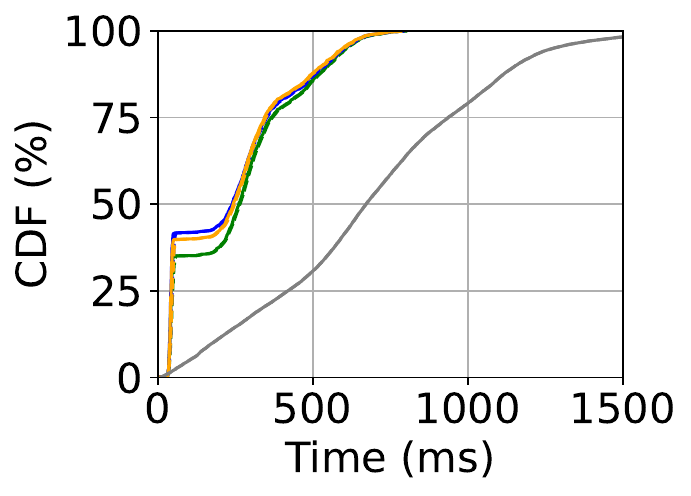} 
    \caption{\scriptsize Seeding (from start)}
    \label{exp:prototype:times:time_to_seeding}
    \end{subfigure}
    \begin{subfigure}{0.22\textwidth} \centering
    \includegraphics[scale=\plotscale]{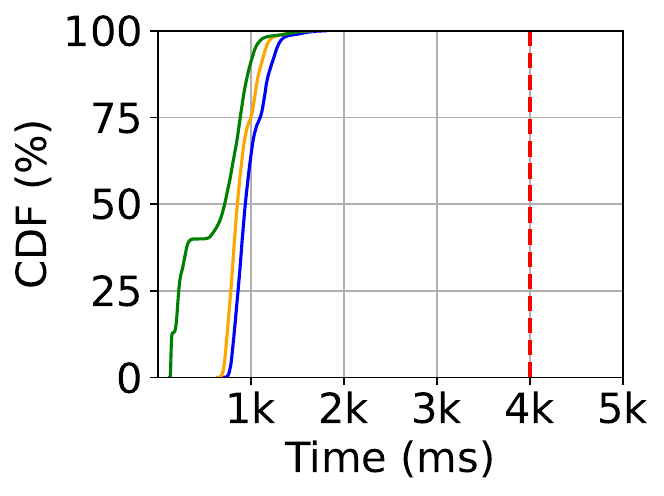}
        \caption{\scriptsize Consolidation (from seeding)}
    \label{exp:prototype:times:time_to_consolidation_from_seeding}
    \end{subfigure}
    \begin{subfigure}{0.22\textwidth} \centering
    \includegraphics[scale=\plotscale]{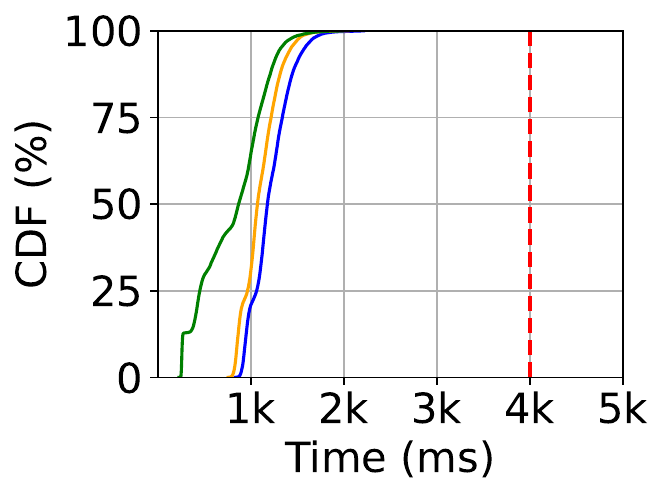}
    \caption{\scriptsize Consolidation (from start)}
    \label{exp:prototype:times:time_to_consolidation_from_start}
    \end{subfigure}
    \begin{subfigure}{0.18\textwidth} \centering
    \includegraphics[scale=\plotscale]{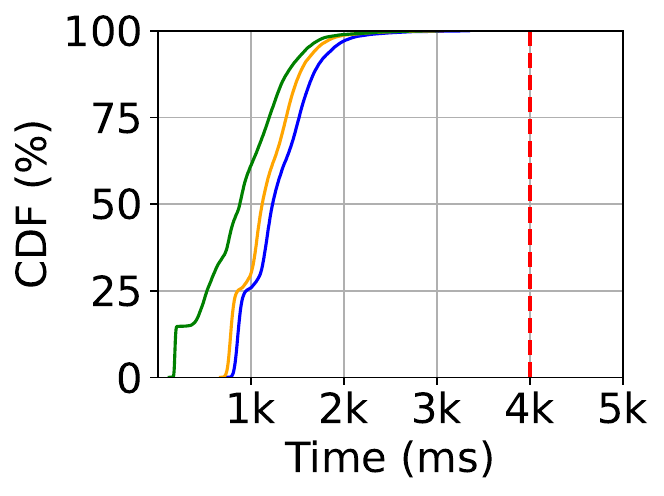}
    \caption{\scriptsize Sampling (from start)}
    \label{exp:prototype:times:time_to_sampling}
    \end{subfigure}
    \begin{subfigure}{0.08\textwidth} \centering
    \includegraphics[scale=\plotscale]{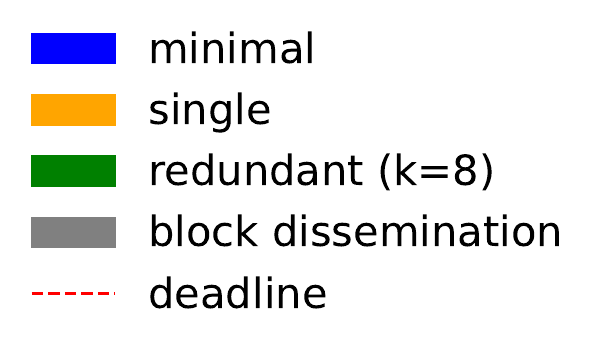}
    \captionsetup{labelformat=empty}
    \caption*{}  
    \label{exp:prototype:times:legend}
    \end{subfigure}

    \vspacebeforecaption
    \caption{Distribution of times for the three phases of \sysname across all nodes, for the three seeding strategies.
    All times are from the start of the slot, except for \Cref{exp:prototype:times:time_to_consolidation_from_seeding} where time is counted from the reception of the seed cells (as shown by \Cref{exp:prototype:times:time_to_seeding}).  }
    \vspaceaftercaption
    \label{exp:prototype:times}
    \end{figure*}

\sysname is implemented in Go, extending \texttt{libp2p}~\cite{libp2p}, the network stack of the Ethereum Geth client~\cite{geth}.
Block dissemination relies on \texttt{libp2p}'s GossipSub implementation.

We aim to evaluate the \sysname prototype in a real-world environment and verify its scalability in large networks, up to 20,000 nodes. 
Achieving both objectives with the \sysname prototype would require a prohibitive amount of resources.
We thus opt for a hybrid approach.
We deploy 1,000 instances of \sysname in a cluster, emulating representative WAN latencies.
To evaluate \sysname up to 20,000 nodes, we use a simulator, whose accuracy is validated against deployment results.

\vspacebeforesubsection
\subsection{Prototype deployments}
\vspaceaftersubsection
\label{sec:evaluation:prototype}

We run 1,000 \sysname instances on a cluster of 80 servers, each equipped with an 18-core Intel Xeon Gold 5220 CPU and 96~GB of RAM.
This level of consolidation (13 instances per server) was selected through careful load testing to avoid CPU contention and increased latencies compared to non-consolidated deployments.

\mypara{Network emulation}
We use network emulation using \texttt{tc} to reproduce WAN settings. 
There is no publicly available data on node-to-node latencies in the Ethereum network.
However, a recent large-scale measurement campaign~\cite{ipfs_latencies} has collected all-pair latencies in IPFS~\cite{benet2014ipfs}, a planetary-scale storage system that shares the scale and decentralization objectives of Ethereum.
We use this trace for our network emulation.
Round-trip latencies range from 8~ms to 438~ms with an average of 64~ms.
The topology contains 10,000 vertices, to which we assign nodes randomly.
We limit each node connection to 25~Mbps.
We deploy a builder as a dedicated server, with a connection capped to 10~Gbps, assigning it to a vertex in the topology randomly selected among the 20\% with the best average latency to all other nodes, i.e., nodes likely deployed in a cloud.
UDP communication in the cluster is subject to a packet loss rate of 3\%, according to our observations.

\mypara{Evaluation metrics}
Our primary metric of interest is the distribution of completion times for \sysname's three phases, from the moment the builder is selected.
The \emph{time to seeding} is when a node has received its initial seed data.
\emph{Time to consolidation} and \emph{time to sampling} refer to the periods when a node has received (or can reconstruct) its assigned eight rows and columns, and its 73 random cells, respectively.
Additionally, we monitor the bandwidth costs and the number of messages for all nodes and the builder.
We consider a fault-free scenario in this section, where all nodes participate in the protocol and have a complete view of the system (i.e., $\forall n \in N, V_n = N$).
For all experiments, we present distributions over 10 slots (i.e., 10 cycles of seeding, consolidation, and sampling).

\mypara{Phases timing}
\Cref{exp:prototype:times} presents the distributions of times to seeding, consolidation, and sampling.
We consider the three seeding strategies of \Cref{sec:pandas_phases:seeding}: \emph{minimal}, \emph{single}, and \emph{redundant} with $k=8$.
Only solid lines are of interest in this section; dashed ones represent simulator results that we will discuss in the next section. \er{CHECK that there are the dashed lines in \Cref{exp:prototype:times}.}
We illustrate the distribution of the reception time of the block via a global GossipSub channel (initiated by a randomly chosen node serving as the proposer), for comparison purposes, in \Cref{exp:prototype:times:time_to_seeding}.

We observe that the time to seeding is similar for the three strategies, as our builder's available bandwidth is not a bottleneck (the amount of data sent out is 36.6~MB, 149~MB, and 1,208~MB, respectively, for the three strategies).
We observe an impact on the tail of the distributions: the maximum time to seeding is 700, 819, or 936~ms, respectively, for the three strategies (99$^\text{th}$ percentiles, or P99, are 698, 705, and 715~ms).
The ``step'' around 64~ms corresponds to nodes assigned to well-connected vertices in the emulated topology, which are typically nodes deployed in the same cloud and/or region as the builder.
Overall, all nodes receive their seed cells before the end of the first second of the slot.

We present the time to consolidation both from the reception of the seed data by a node (\Cref{exp:prototype:times:time_to_consolidation_from_seeding}) and from the beginning of the slot (\Cref{exp:prototype:times:time_to_consolidation_from_start}).
The builder provides the consolidation boosting map to the nodes.
We can observe the impact of the builder's seeding strategy.
The minimal strategy results in a consolidation taking up to 2,2213~ms (P99=1,756~ms) from the reception of the seed data, and the single strategy has a maximum time of 2,046~ms (P99=1,595~ms).
In contrast, the redundant strategy reduces this time to 1,985~ms (P99=1,558~ms).
Median times to consolidation for the minimal, single, and redundant strategies (from the beginning of the slot) are 1,178~ms, 1,072~ms, and 869~ms, respectively. 

The time to sampling distribution, our primary metric of interest, is given by \Cref{exp:prototype:times:time_to_sampling}.
This distribution depends on the builder's seeding strategy, which impacts the time to consolidation.
The minimal strategy results in a maximum of 3,341~ms (P99=2,303~ms); still, 100\% of the nodes fetch their samples by the deadline.
The single strategy also meets the deadline, with a maximum delay of 3,062~ms, (P99=2,068~ms).
Finally, the redundant strategy matches the deadline safely for all nodes, with a maximum of 3,009~ms (P99=2,020~ms).
The median times to sampling are, respectively, 1,235~ms, 1,122~ms, and 882~ms.
The reduction in sampling times with increased availability of seed cells (via increased redundancy) is due to reduced contention on peer bandwidth, which in turn speeds up the fetching operation.
We observe, however, that if the block dissemination latency (\Cref{exp:prototype:times:time_to_seeding}) were to be added to these times, meeting the 4~s deadline would be at risk for many nodes, even with the redundant strategy.
This confirms our claim that DAS must start \textit{concurrently} to block dissemination if we are to integrate it with consensus under the tight fork-choice rule. 

\begin{figure}[t]
    \centering
    \begin{subfigure}{0.45\textwidth}
        \centering
        \includegraphics[scale=\plotscale]{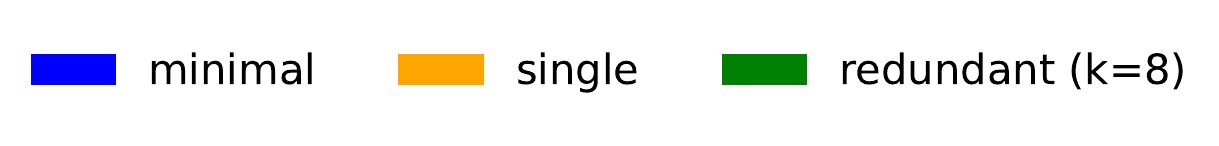}
        \captionsetup{labelformat=empty}  
        \caption*{}  
    \end{subfigure}
    \vspace{-2em}

    \begin{subfigure}[b]{.49\linewidth}
        \centering
    \includegraphics[scale=\plotscale]{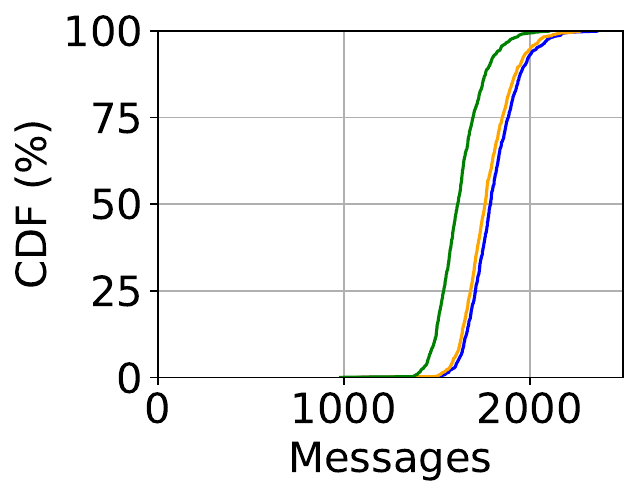}
    \caption*{}  
    \end{subfigure}
    \hfill
    \begin{subfigure}[b]{.49\linewidth}
        \centering
        \includegraphics[scale=\plotscale]{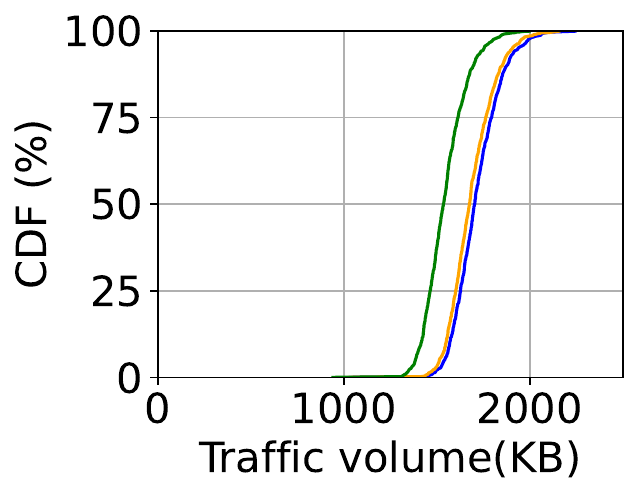}
        \caption*{}  
    \end{subfigure}

    \vspacebeforecaption
    \caption{Distribution of messages and traffic volume for fetching across nodes, for different seeding strategies.}
    \vspaceaftercaption
    \label{exp:prototype:messages}
\end{figure}

\mypara{Bandwidth consumption}
\Cref{exp:prototype:messages} presents the distribution of the number of messages used by nodes in the fetching phase, and the corresponding traffic volume, summed for both directions.
The redundant seeding strategy results in fewer messages exchanged between nodes and, as a result, lower bandwidth requirements.
This is because more nodes already hold the requested cells, which reduces the need for retries (i.e., fewer rounds) during consolidation and sampling.
Even with the single seeding strategy, which only marginally differs from the minimal one, the requirements are far below the Ethereum recommendations of 25~Mbps; the maximum traffic volumes are 2.26, 2, and 1.99~MB for the three strategies.

\begin{table}[t]
\footnotesize
\newcolumntype{P}{>{\raggedright\arraybackslash} m{0.32\linewidth} }
\newcolumntype{Q}{>{\raggedright\arraybackslash} m{0.115\linewidth} }
\renewcommand{\arraystretch}{1}

\begin{tabular}{PQQQQ}
\toprule
\textbf{Round} & \textbf{1} & \textbf{2} & \textbf{3} & \textbf{4} \\
\midrule
Messages sent                 & $341 {\scriptstyle\,\pm\, 20}$    & $261 {\scriptstyle\,\pm\, 58}$  & $185 {\scriptstyle\,\pm\, 35}$   & $113 {\scriptstyle\,\pm\, 22}$ \\
Cells requested               & $4174 {\scriptstyle\,\pm\, 100}$  & $2426 {\scriptstyle\,\pm\, 96}$ & $923 {\scriptstyle\,\pm\, 63}$   & $294 {\scriptstyle\,\pm\, 40}$ \\
Replies received in round     & $228 {\scriptstyle\,\pm\, 22}$    & $143 {\scriptstyle\,\pm\, 14}$  & $120 {\scriptstyle\,\pm\, 20}$   & $69 {\scriptstyle\,\pm\, 25}$ \\
Replies received after round  & $107 {\scriptstyle\,\pm\, 39}$    & $114 {\scriptstyle\,\pm\, 25}$  & $56 {\scriptstyle\,\pm\, 20}$    & $61 {\scriptstyle\,\pm\, 3}$ \\
Cells received in round       & $2420 {\scriptstyle\,\pm\, 180}$  & $949 {\scriptstyle\,\pm\, 170}$ & $535 {\scriptstyle\,\pm\, 82}$   & $191 {\scriptstyle\,\pm\, 22}$ \\
Cells received after round    & $1128 {\scriptstyle\,\pm\, 113}$  & $1478 {\scriptstyle\,\pm\, 91}$ & $383 {\scriptstyle\,\pm\, 52}$   & $23 {\scriptstyle\,\pm\, 8}$ \\
Received cells duplicates     & $0 {\scriptstyle\,\pm\, 0}$       & $187 {\scriptstyle\,\pm\, 42}$  & $142 {\scriptstyle\,\pm\, 29}$   & $64 {\scriptstyle\,\pm\, 12}$ \\
Cells reconstructed           & $615 {\scriptstyle\,\pm\, 126}$   & $566 {\scriptstyle\,\pm\, 90}$  & $86 {\scriptstyle\,\pm\, 29}$    & $32 {\scriptstyle\,\pm\, 17}$ \\
Cumulative coverage of $F$    & 56\%                             & 81\%                           & 96\%                            & 99\%         \\

\bottomrule
\end{tabular}

\caption{Fetching algorithm performance in successive rounds (values averaged over all nodes, $\pm$ is the standard deviation).
\vspaceaftercaption
}
\label{tab:evaluation:fetching_per_round}
\end{table}


\mypara{Fetching analysis}
\Cref{tab:evaluation:fetching_per_round} presents an analysis of the progress of fetching for the first four rounds.
All values discussed in this paragraph are averages over the 1,000 nodes, together with the standard deviation.
We use the redundant seeding strategy, and nodes receive 2420 cells ($\pm$ 180).
Starting from 4,174, the number of requested cells decreases as the coverage of $F$ (\ie set of cells to fetch) increases, either through reception or reconstruction (e.g., 615 reconstructed cells in the first round).
We distinguish between replies received \emph{in} a round $i$, i.e., before the timeout $t_i$ expires, and after.
The latter case leads to redundant requests but illustrates the tradeoff between caution and eagerness \Cref{alg:fetching} implements.
A majority of requests result in replies before the timeout, and a majority of cells are received on time in the round.
Receptions after the round generally occur with a significant delay; adjusting timeouts to account for such tail latencies leads to lower success rates.
While \Cref{tab:evaluation:fetching_per_round} only shows the first four rounds---after which 99\% of the nodes have completed fetching---the process requires up to 6 rounds for the slowest nodes (P90=3, P99=4).


\begin{figure}[t]
    \centering
    \begin{subfigure}{0.45\textwidth}
        \centering
        \includegraphics[scale=\plotscale]{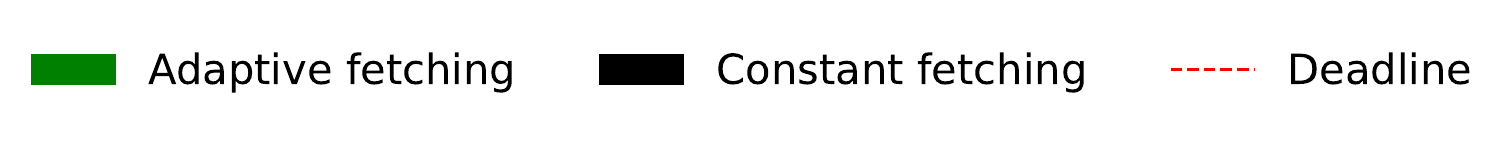}
        \captionsetup{labelformat=empty}  
        \caption*{}  
    \end{subfigure}
    \vspace{-2em}

    \begin{subfigure}[b]{.49\linewidth}
        \centering
        \includegraphics[scale=\plotscale]{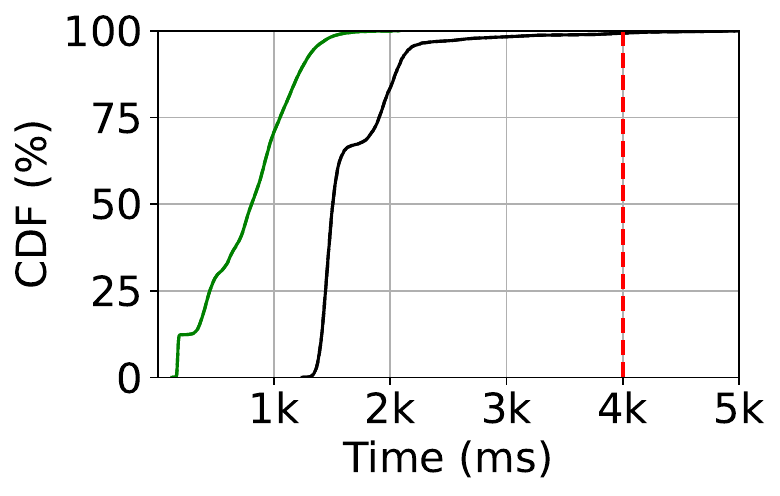} 
        \caption{Time to sampling}
        \label{fig:randomSampling:time}
    \end{subfigure}
    \hfill
    \begin{subfigure}[b]{.49\linewidth}
        \centering
        \includegraphics[scale=\plotscale]{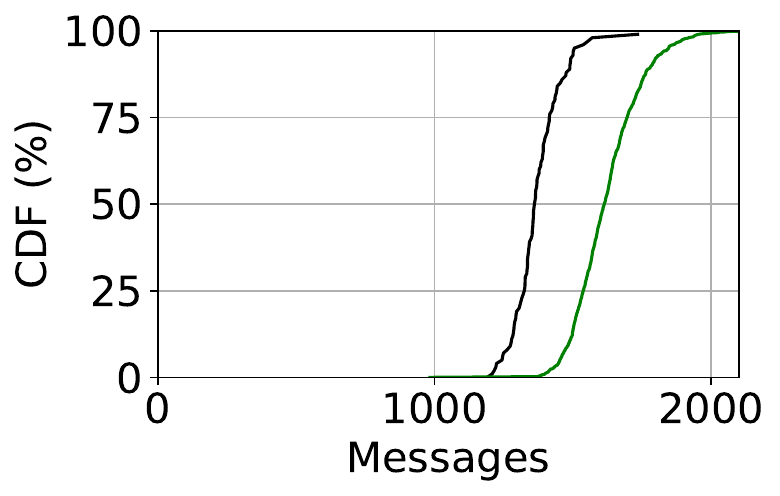}
        \caption{Messages in and out}
        \label{fig:randomSampling:messages}
    \end{subfigure}
    \vspacebeforecaption
    \caption{Comparison of the performance of adaptive fetching, as used by \sysname, and a non-adaptive approach.}
    \vspaceaftercaption
    \label{fig:randomSampling}
\end{figure}

\mypara{Impact of adaptive fetching}
We evaluate in \Cref{fig:randomSampling} the impact of adaptive fetching.
We consider the redundant seeding policy, i.e., the green distribution in \Cref{fig:randomSampling:time} is the same as in \Cref{exp:prototype:times:time_to_sampling}.
For comparison, we employ a \emph{constant} fetching strategy, which utilizes a fixed timeout for all rounds ($t=400~\text{ms}$) and a fixed redundancy ($k=1$), as represented in black.
The constant strategy uses fewer messages, as it asks only a minimum of one node for each missing cell in each round, and leaves more time for nodes to respond.
However, it drastically impacts the time to sampling, resulting in a maximum of 4,129~ms (P99=3,513~ms, median=1,546~ms), and some nodes miss the deadline.
This illustrates the interest of dynamically adapting aggressiveness and redundancy to cope with the tight time constraints imposed by the tight fork-choice rule.

\begin{figure}[t]
    \centering
    \begin{subfigure}[b]{.45\linewidth}
        \centering
        \includegraphics[scale=\plotscale]{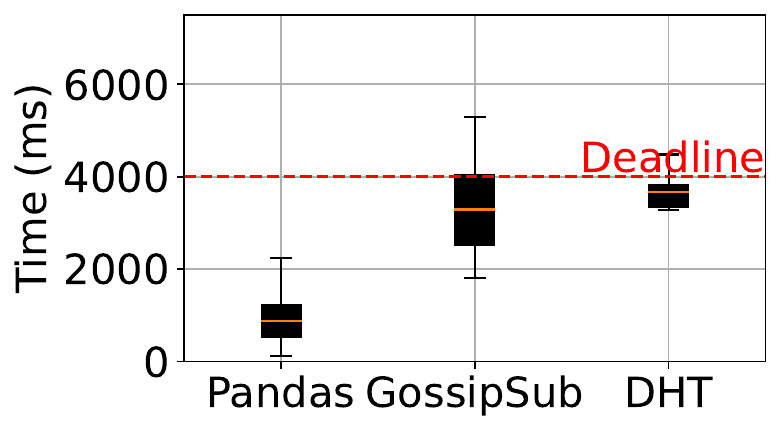} 
        \caption{Time to sampling}
        \label{fig:testbed2:time}
    \end{subfigure}
    \begin{subfigure}[b]{.45\linewidth}
        \centering
        \includegraphics[scale=\plotscale]{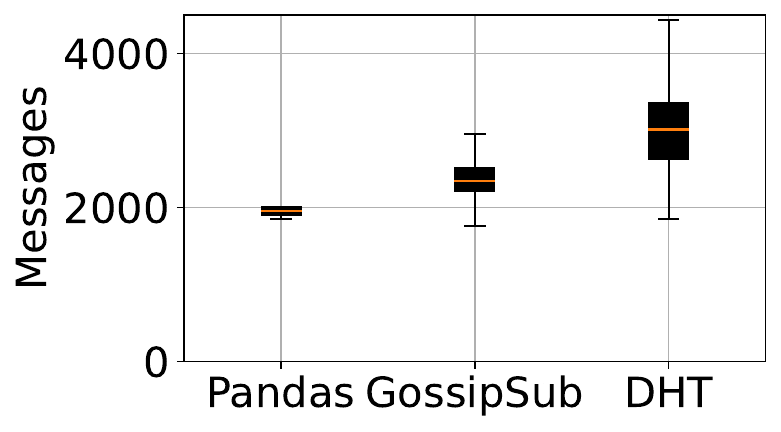}
        \caption{Messages}
        \label{fig:testbed2:messages}
    \end{subfigure}
    \vspacebeforecaption
    \caption{Distribution of time to sampling and messages compared to baselines based on GossipSub and the Kademlia DHT.
    \vspaceaftercaption
    }
    \label{fig:testbed2}
\end{figure}

\mypara{Comparison to alternative proposals}
We finally compare our approach to two alternative methods based on the use of existing peer-to-peer protocols available in \texttt{libp2p}.

Some proposals~\cite{subnetdas,fulldas,peerdas}, that we will further discuss in \Cref{sec:related}, suggest using GossipSub~\cite{gossipsub} for the dissemination of rows and columns (i.e., our seeding phase), but typically do not detail how random sampling should happen.
We instantiate this idea by having all nodes subscribe to GossipSub channels corresponding to each unit of custody---that is, each group of eight rows and eight columns\er{note that this is different than what we do in \sysname, and probably at the advantage of the GossipSub baseline. If we were to assign rows and columns with $\sigma$ in packs of 8 to the same nodes we would get probably better results.}---they are assigned by $\sigma$. We disable explicit consolidation, but instead rely on GossipSub’s gossiping within each channel to disseminate the assigned cells, and use the same sampling phase as in \sysname.
Therefore, the main difference is that the dissemination of seed cells occurs through peer-to-peer gossip within each channel, rather than through direct seeding by the builder as in \sysname.
In this 1,000-node network, each GossipSub channel involves approximately $\sim16$ nodes assigned to the corresponding unit of custody.
The builder sends $k=8$ copies of each unit of custody to the nodes in the corresponding GossipSub channel, which is configured with the default fanout of eight peers.
As a result, the builder's outgoing traffic volume is the same as in the redundant seeding strategy of \sysname, i.e., eight times the total blob size.

Another proposed approach~\cite{cortesgoicoechea2024scalability} is to use the Kademlia DHT~\cite{maymounkov2002kademlia} for storing and retrieving cells using multi-hop routing.
We implement it by mapping rows and columns to one dimension and splitting it into \emph{parcels} of 64 adjacent cells.
Parcels are then stored in the DHT by the builder using the $\algvar{put(key)}$ operation.
To ensure a fair comparison with \sysname and the GossipSub baseline, the builder performs eight $\algvar{put(key)}$ operations per parcel, storing it at each of the eight closest peers to the hash of the parcel’s contents---therefore, the builder uses the same total bandwidth as in the other approaches.
Nodes are responsible for the range of keys (and, therefore, parcels) assigned by the DHT, and we disallow consolidation.
Sampling uses $\algvar{get(key)}$ operations to fetch necessary parcels.


\Cref{fig:testbed2} shows the distribution of time to sampling for \sysname using the redundant seeding strategy ($k=8$) and for the two baselines, as well as the distribution of the number of messages.
With 1,000 nodes, 24\% of GossipSub nodes and 17\% of DHT nodes fail to complete sampling within the 4~s deadline. The average sampling delay for GossipSub nodes is 3,660~ms (P99=3342~ms), while \sysname nodes complete sampling significantly faster, i.e., on average in 882~ms (P99=1935~ms), with all nodes completing well within the 4~s deadline.
In terms of messaging, the DHT and GossipSub baselines incur significantly higher overhead in the number of messages compared to \sysname.
On average, \sysname, GossipSub, and DHT nodes send 1,613, 2,370, and 3,021 messages.
For the DHT baseline, the messaging overhead of storing and retrieving parcels is especially high due to multi-hop routing (i.e., DHT traversal).


%

\vspacebeforesubsection
\subsection{Large-scale simulations}
\vspaceaftersubsection
\label{sec:evaluation:simulations}

In addition to the prototype deployment detailed in the previous subsection, we implement \sysname protocols in PeerSim~\cite{p2p09-peersim}, a Java simulator for large-scale evaluation of peer-to-peer systems.
This implementation closely follows the one over \texttt{libp2p}.
We also implement the two baselines detailed above.

We simulate the same latency trace as for the deployment.\footnote{When using more than 10,000 nodes, we reuse vertices randomly for the assignment.}
We also enforce a fixed 3\% loss rate for UDP packets as experienced in the testbed.
When running on a server with 256~GB of memory, the simulator can scale up to 20,000 nodes.

\mypara{Simulator validation}
Before considering larger scales, we validate that the simulator results match those of the deployments.
In all plots of \Cref{sec:evaluation:prototype}, dashed lines report the results obtained with 1,000 simulated nodes.
In all cases, the two lines are (almost) indistinguishable.
Our evaluations at smaller scales (not shown) have the same property.
The validation of simulation results at moderate scales gives us confidence in the simulator's ability to provide accurate results at higher scales.

\begin{figure}[t]
    \centering
    
    \begin{subfigure}{0.45\textwidth}
        \centering
        \includegraphics[scale=\plotscale]{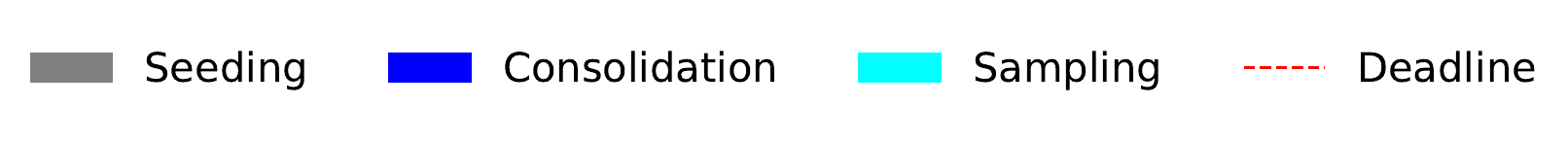}
        \captionsetup{labelformat=empty}  
        \caption*{}  
    \end{subfigure}
    \vspace{-2em}
    
    \begin{subfigure}[b]{.33\linewidth}
        \centering
        \includegraphics[scale=\plotscale]{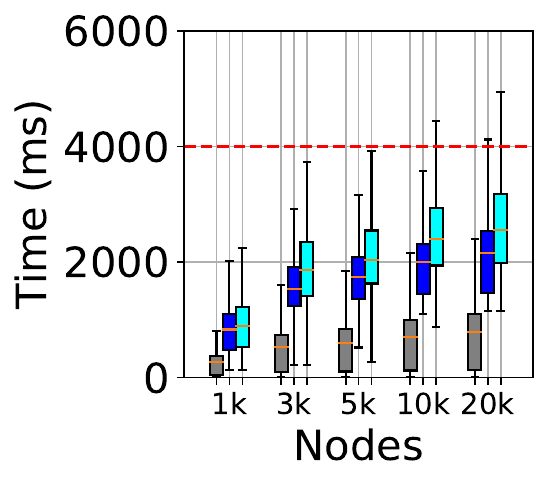} 
        \caption{\scriptsize Time}
        \label{fig:sim:pandas:time}
    \end{subfigure}
    \hfill
    \begin{subfigure}[b]{.3\linewidth}
        \centering
        \includegraphics[scale=\plotscale]{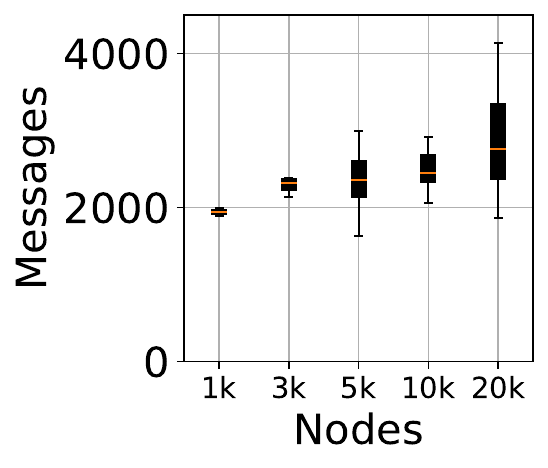}
        \caption{\scriptsize Fetching messages}
        \label{fig:sim:pandas:messages}
    \end{subfigure}
    \hfill
    \begin{subfigure}[b]{.33\linewidth}
        \centering
        \includegraphics[scale=\plotscale]{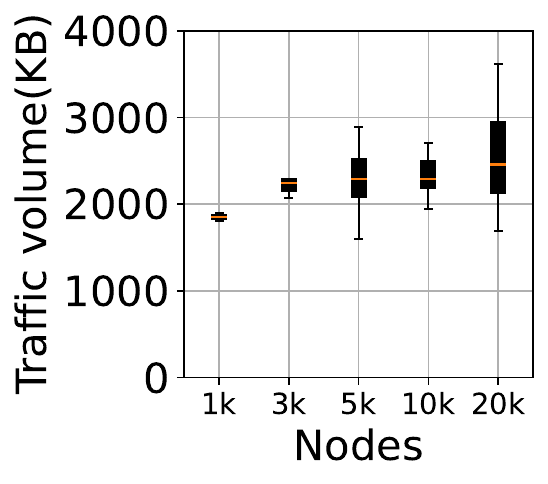}
        \caption{\scriptsize Fetching bandwidth}
        \label{fig:sim:pandas:bandwidth}
    \end{subfigure}
    \vspacebeforecaption
    \caption{Simulation of seeding, consolidation, and sampling times for \sysname with a various number of nodes.
    \vspaceaftercaption
    }
    \label{fig:sim:pandas}
\end{figure}


\mypara{Scaling}
We first investigate \sysname's scalability with 1,000 to 10,000 nodes and up to 20,000 nodes.
\Cref{fig:sim:pandas} presents the distribution of times to seeding, consolidation, and sampling using the redundant seeding strategy (\Cref{fig:sim:pandas:time}) and the corresponding messages (\Cref{fig:sim:pandas:messages}) and bandwidth (\Cref{fig:sim:pandas:bandwidth}).
With 10,000 nodes, the current scale of the Ethereum network~\cite{etherscan_may2025}, all nodes successfully sample before the 4~s deadline.
With 20,000 nodes, 10\% fail to meet the deadline, mapping to nodes with poor simulated connectivity (connected from remote area of the geo-distributed network).
We identify the scattering of seed data and the cost of consolidation as the primary reasons, as nodes have to contact more peers to collect their rows and samples, and take more time before being able to answer sampling requests.
Nodes located in clouds do not suffer from significantly higher times, highlighting the need to host validator-hosting nodes in well-connected infrastructure. 

The impact of the increasing scattering of seed cells with larger network sizes is also reflected in \Cref{fig:sim:pandas:messages} and \Cref{fig:sim:pandas:bandwidth}, which show the number of messages and traffic volumes for fetching cells during consolidation and sampling. The average number of messages per node for networks of 1K, 3K, 5K, 10K, and 20K nodes is 1,956, 2,231, 2,247, 2,291, and 2,443, respectively. 
The corresponding peak traffic volumes are 1.9, 2.1, 2.2, 2.2, and 2.4~MB.
We observe that even in the most demanding scenario with 20K nodes, the maximum traffic volume is transmitted in approximately 2.2 seconds, keeping the average bandwidth requirement well within the 25~Mbps target.

%

\begin{figure}[t]
    \centering
    \begin{subfigure}{0.45\textwidth}
        \centering
        \includegraphics[scale=\plotscale]{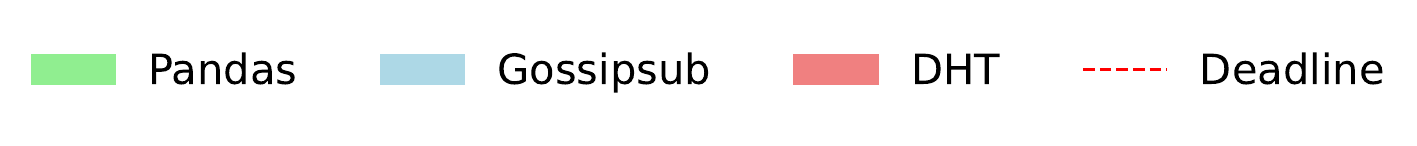}
        \captionsetup{labelformat=empty}  
        \caption*{}  
    \end{subfigure}
    \vspace{-2em}

    \begin{subfigure}[b]{.3\linewidth}
        \centering
        \includegraphics[scale=\plotscale]{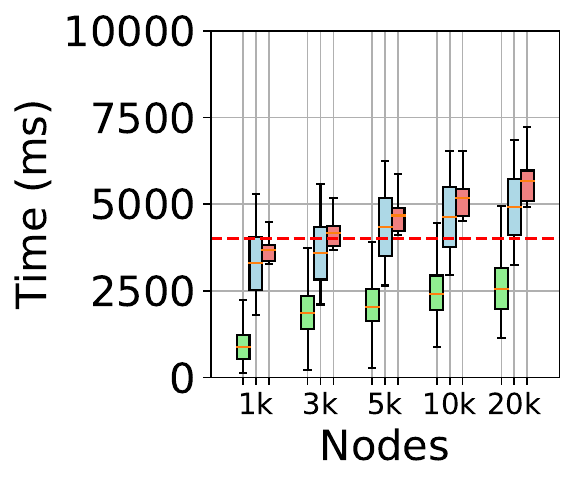} 
        \caption{Time to sampling}
        \label{fig:sim:compare:time}
    \end{subfigure}
    \hfill
    \begin{subfigure}[b]{.3\linewidth}
        \centering
        \includegraphics[scale=\plotscale]{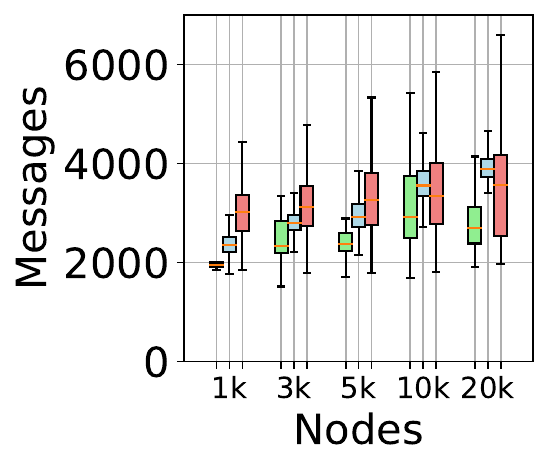}
        \caption{Messages}
        \label{fig:sim:compare:messages}
    \end{subfigure}
    \hfill
    \begin{subfigure}[b]{.3\linewidth}
        \centering
        \includegraphics[scale=\plotscale]{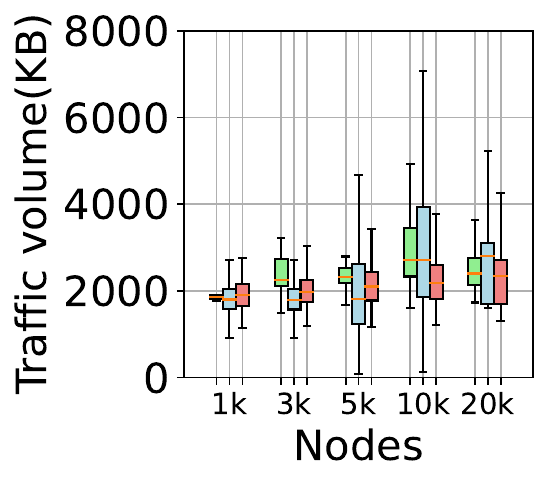}
        \caption{Bandwidth}
        \label{fig:sim:compare:bandwidth}
    \end{subfigure}
    \vspacebeforecaption
    \caption{Simulation of blob dissemination time for \sysname and the two baselines, for various number of nodes.
    \vspaceaftercaption
    }
    \label{fig:sim:compare}
\end{figure}

\mypara{Comparison to baselines}
\Cref{fig:sim:compare} compares \sysname to the two baselines in scales up to 20,000 nodes.
Results for 1,000 nodes are consistent with the ones for testbed deployment reported in \Cref{fig:testbed2}.
While the GossipSub-based baseline meets the deadline for a majority of nodes with 1,000 nodes, it fails to do so starting with 5,000 nodes.
However, it plateaus for higher node counts, as GossipSub topics become more efficient with a higher number of participants.
The DHT-based baseline is unable to meet deadlines for most nodes at all scales and shows linearly increasing times to sampling for increased system sizes.
For both systems, the gap to \sysname in terms of time-to-sampling latency widens as the system size grows.
The number of messages is also significantly higher for the baselines than for \sysname, with important variability for the GossipSub-based baseline as the system size increases.
\er{the values for bandwidth in \Cref{fig:sim:compare:bandwidth} are inconsistent with \Cref{exp:prototype:messages}.}

\begin{figure}[t]

    \begin{subfigure}{0.45\textwidth}
        \centering
        \includegraphics[scale=\plotscale]{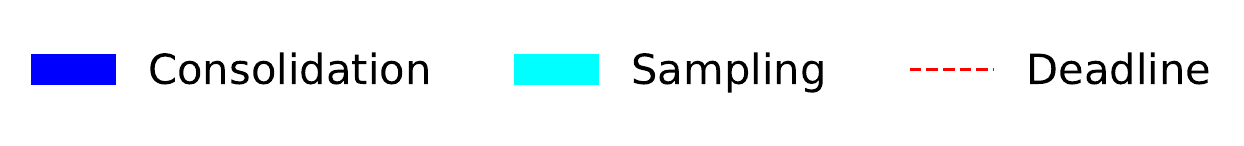}
        \captionsetup{labelformat=empty}  
        \caption*{}  
    \end{subfigure}
    \vspace{-2em}

    \begin{subfigure}[b]{.45\linewidth}
        \centering
        \includegraphics[scale=\plotscale]{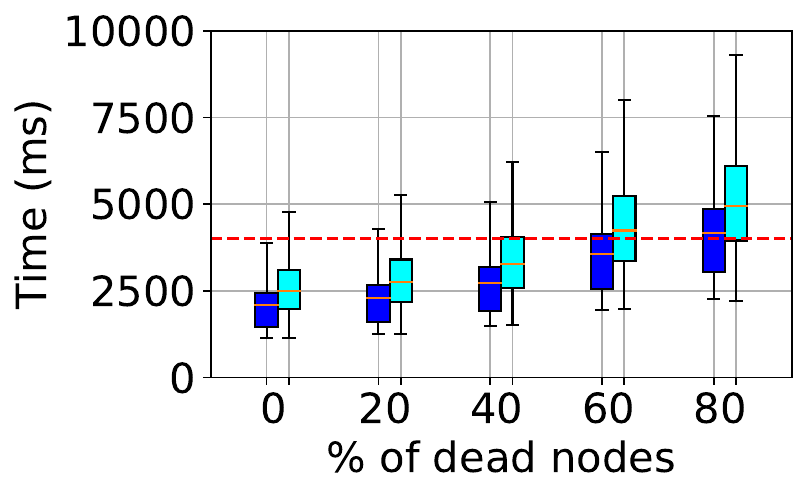} 
        \caption{Dead nodes}
        \label{fig:unperfect:unperfect}
    \end{subfigure}
    \hfill
    \begin{subfigure}[b]{.45\linewidth}
        \centering
        \includegraphics[scale=\plotscale]{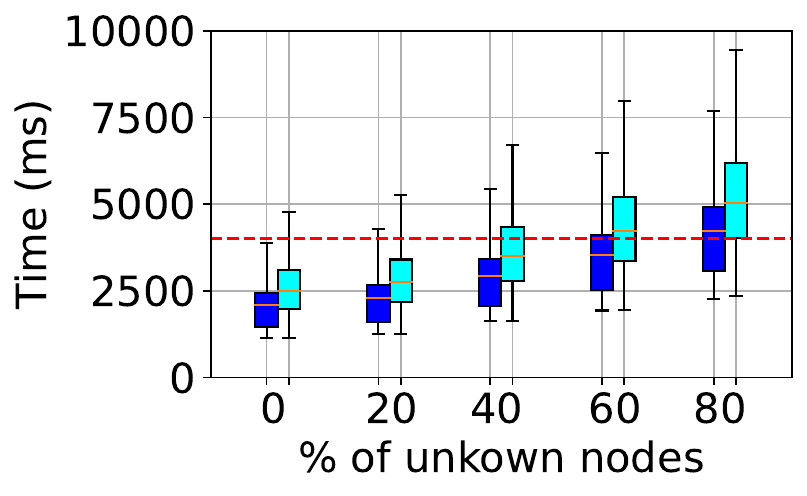}
        \caption{Out-of-view nodes}
        \label{fig:unperfect:dead}
    \end{subfigure}
    \vspacebeforecaption
    \caption{Simulation of time to consolidation and time to sampling for increasing number of dead nodes and out-of-view nodes, in a 10.000-node network.
    \vspaceaftercaption
    }
    \label{fig:unperfect}
\end{figure}

\mypara{Behavior under faults}
\er{TODO comment \Cref{fig:unperfect}} \onur{Added the below text as a placeholder}
We evaluate \sysname's robustness under two types of faults: \emph{dead nodes} and \emph{out-of-view nodes}.
In the \emph{dead nodes} scenario, a fraction of nodes is assumed to have crashed and do not respond to any messages. 
The builder and the remaining correct nodes are unaware of these failures.
Therefore, the builder seeds data to all nodes, including the dead ones, and includes them in consolidation boost maps.
As a result, some seed cells are lost, and correct nodes may attempt to contact dead nodes during fetching which will lead to timeouts and retries.
On the other hand, in the \emph{out-of-view nodes} scenario, all nodes are correct and receive their assigned seed cells from the builder.
However, each node only has an incomplete view of the network, and these views are not consistent.
For example, if 20\% of nodes are out-of-view, each node is only aware of a randomly chosen 80\% of the full node set. 
This affects both consolidation and sampling, as requests may fail due to the sender lacking the knowledge of a suitable peer.

In \Cref{fig:unperfect}, we vary the proportion of dead or out-of-view nodes from 0\% to 80\% (in 20\% increments) and measure the impact on both time to consolidation and sampling.
The network size is 10,000 nodes.
We observe that for 0\%, 20\%, 40\%, 60\%, and 80\% of \emph{dead nodes}, 92\%, 83\%, 74\%, 45\%, and 27\% of nodes complete sampling within the 4-second deadline, respectively.
In the case of \emph{out-of-view nodes}, for 0\%, 20\%, 40\%, 60\%, and 80\% of nodes being out of view, 92\%, 83\%, 67\%, 47\%, and 25\% of nodes complete sampling within the deadline, respectively.

In both scenarios, we observe that beyond 50\% dead or out-of-view nodes, over half of the correct nodes fail to meet the deadline which would prevent consensus from being reached and causing the blockchain to stall.
These scenarios are, however, unlikely in practice and would impact other key mechanisms, such as block dissemination, preventing consensus to succeed anyway.

\mypara{Summary}
Our evaluation using a prototype deployment and large-scale simulations of \sysname and two baselines confirmes our claims.
\sysname supports DAS within 4~s for all nodes (up to 10,000) and the vast majority of them (for 20,000) and enable the tight fork-choice rule (\textbf{C1} and \textbf{C4}).
The bandwidth requirements for builders and nodes are below Ethereum recommendations and compatible with its decentralization objectives under the PBS principles (\textbf{C2} and \textbf{C4}).
The evaluation of fault scenarios show that \sysname supports these claims with a large fraction of failed or out-of-view nodes (\textbf{C3} and \textbf{C4}).
In contrast, the two baselines fail to meet these criteria in particular as the system size increases (\textbf{C5}).

%% file: sections/discussion.tex
\vspacebeforesection
\section{Discussion}
\vspaceaftersection
\label{sec:discussion}

We discuss \sysname and classical concerns in decentralized systems.


\mypara{Impact of Sybils}
%
A node in Ethereum, and thus \sysname, does not have to support validators to participate in peer-to-peer interactions, e.g., block dissemination.
This opens possibilities for \emph{Sybil} attacks, where an attacker operates multiple nodes to bias the system operation.

Sybils can perform \emph{general} attacks, where they join the DHT and GossipSub channels and stop answering queries or forwarding data, disrupting the system merely by their overwhelming presence.
Our evaluation shows that \sysname is robust against many \michal{up to X\%; or even for X\%} nodes that ignore sampling and consolidation requests, provided the builder uses sufficient redundancy in its seeding strategy.
Proposals for increasing 
IP diversity in Ethereum's discovery mechanisms~\cite{krol2024disc} could strengthen this robustness.

A \emph{targeted} use of Sybils consists of carefully placing them in the peer-to-peer network to prevent specific nodes from interacting with it (an \emph{Eclipse} attack) or to censor specific information~\cite{henningsen2019eclipsing,wust2016ethereum,marcus2018low,sridhar2024content}.
\sysname makes the network fully connected and randomized exchanges, making Eclipse attacks irrelevant.
This contrasts with designs based on GossipSub trees, where an attacker could position its Sybils as the first neighbors of the builder and disrupt the early dissemination of blob data. 
Another targeted attack scenario targets specific \emph{content}. 
In \sysname, disrupting the sharing of specific blob data
would require
    (1)~knowing before blob data dissemination which cell will contain such data
    and
    (2)~positioning Sybil nodes in the network to make the corresponding row and column difficult to reconstruct (i.e., disallow fetching half of its cells).
Condition (1) does not hold as the cell location is known only upon reception of the block.
Condition (2) would require the attacker to generate enough identities to control the corresponding row and column, which is highly improbable considering that $\sigma$ changes unpredictably every 6.5 minutes, less time than what ENR crawling requires.



\mypara{Limiting openness}
As one of \sysname goals is to avoid modification to Ethereum other mechanisms, it follows its open-network design.
An alternative design could limit participation to validator-holding nodes and restrict other nodes to being only observers.
This would drastically reduce any potential risk associated with Sybil attacks, as an attacker can only generate one identity for every 32 ETH they hold. 
It would, however, limit decentralization by switching to a semi-permissioned system, where only stakeholders can participate.
\emph{Proof-of-validator}~\cite{proof_of_validator} is an anonymous credential scheme based on zero-knowledge proofs (ZKP).
It could enable this limitation if integrated with node discovery, i.e., the crawl of the DHT for ENRs. 
This approach would come at a significant complexity cost, even for an observer (i.e., gathering the list of validators and verifying a ZKP for every crawled ENR).


\mypara{Handling free riders}
Nodes using the system while contributing minimal resources, or free riders, are unavoidable in decentralized systems~\cite{ihle2023incentive}.
In Ethereum, block and blob building, proposal, and validation are incentivized through monetary rewards; however, there is no incentive for interactions within the peer-to-peer network (e.g., for correctly answering ENR discovery requests in the DHT).
Similarly, \sysname does not have incentives for nodes to participate in the consolidation, sampling, and hosting of blob data.
Nodes hosting committee members could even vote for blob availability without sampling, if the expected rewards outweigh bandwidth costs.

Also departing from our objective of no modification to Ethereum, a possible direction to include incentives for DAS operations would be to integrate \emph{proof-of-custody} mechanisms~\cite{proof_of_custody} and the associated slashing mechanisms with consensus.
Proof of custody is a negative incentive in which the builder infrequently inserts in blob data a cryptographic ``bomb'' targeted at a randomly-selected validator.
A node that attests to blob data with a bomb for itself is slashed for a large amount of stake, making the correct behavior of downloading and verifying data more profitable than free riding.
As only nodes holding stake can be targeted, and not observers, such a mechanism would probably depend on proof-of-validator integration.


%% file: sections/related.tex
\vspacebeforesection
\section{Related Work}
\vspaceaftersection
\label{sec:related}

We start by detailing alternative proposals for implementing DAS in Ethereum.
Then, we explore the broader history and foundational literature on data availability. 
Finally, we discuss earlier P2P techniques designed for purpose-specific data dissemination.

\mypara{Alternative DAS Proposals}
The Ethereum community has so far mainly explored gossip-based approaches to support DAS and the Danksharding roadmap, primarily using GossipSub~\cite{gossipsub}, where each row and column is disseminated through a distinct channel with long-lived subscriptions~\cite{peerdas, chaudhuri2024design, fulldas}.
The scalability of this approach remains uncertain due to the large number of channels required and potential security risks associated with long-lived node-to-channel assignments~\cite{kumar2024formal}.
More importantly, the slow, multi-hop propagation of rows and columns resulted in proposals considering the removal of random sampling from the critical path of consensus.
Instead, these approaches adopt a ``lightweight'' custody-based verification, where nodes supporting committee members evaluate the availability of a blob (and vote accordingly) based solely on the successful receipt of their assigned rows and columns~\cite{subnetdas}. 
This approach provides significantly weaker assurances of availability compared to random sampling.
By foregoing independent verification of randomly chosen cells, validators risk being slashed or locked onto an unavailable chain.
This, in turn, could prevent them from forming a minority chain or participating in a manual fork of the canonical chain under social consensus in the event of a malicious majority.

Alternative network-layer mechanisms for DAS have also been evaluated.
A recent study~\cite{cortesgoicoechea2024scalability} highlights the inefficiencies of using the Kademlia DHT~\cite{maymounkov2002kademlia} for DAS, particularly the overhead of seeding cells to nodes involving traversing the DHT.

\para{Data Availability}
The idea of verifying data availability by sampling a block extended with erasure coding was introduced by Al-Bassam \etal~\cite{al2021fraud} and later adopted by LazyLedger~\cite{al2019lazyledger}, since evolving into Celestia.
Celestia employs a \emph{centralized} approach where validators, i.e., highly resourceful ``super'' nodes---retrieve and store the complete blob data, while light clients sample the blob from these super nodes.
Unlike \sysname's collaborative approach, where sampling and storage responsibilities are distributed among participants, Celestia’s design results in overhead and costs that increase linearly with participation and blob size.
Recent work by Nazirkhanova \etal~\cite{nazirkhanova2022information} explores how erasure coding, combined with homomorphic vector commitments, can ensure verifiable data retrieval in rollups while maintaining storage and communication efficiency.

Recent work proposes alternative methods for node sampling in DAS. 
Honeybee~\cite{zhang2024honeybee} focuses on Sybil-resistant peer sampling via verifiable random walks.
Honeybee assumes fetching random cells requires contacting random nodes, whereas \sysname selects random cells and takes advantage of their deterministic assignment to nodes.
Honeybee’s interactive verification at each hop introduces potential latency, making it less suitable for strict timing constraints. 
Sheng \etal~\cite{sheng2021aced} propose an alternative approach where a group of oracle nodes collectively store erasure-coded data blobs and provide access to clients.
Their work focuses on ensuring the integrity and correctness of the coded blob and relies on at least half of the oracle nodes being honest and reliable.
In contrast, \sysname can function even under a supermajority of nodes failing or experiencing omission faults.


 \mypara{Purpose-specific Data Dissemination Networks}
 One-hop communication in overlay networks was proposed in the early days of P2P research~\cite{gupta2003one}.
 The Interplanetary File System (IPFS)~\cite{benet2014ipfs} adopted a hybrid P2P networking approach by combining multi-hop search (i.e., through a Kademlia DHT) and one-hop communication.
 More specifically, each peer accesses a few directly connected peers to perform a one-hop search and retrieve popular content~\cite{de2021accelerating}.
 The DHT is used to discover peers hosting (less popular) content through a slow, multi-hop search process.
 While IPFS supports one-hop retrieval, it is not optimized for the DAS use case, where a large number of directly connected peers must be efficiently utilized to retrieve unpopular content in a timely manner, i.e., chunks of a data blob each hosted by roughly the same number of peers. 

 Beyond IPFS, several studies explore optimizations for blockchain and distributed systems that improve data dissemination efficiency.
 Mercury~\cite{zhou2023mercury} optimizes blockchain transaction dissemination using network coordinates, while Perigee~\cite{mao2020perigee} improves Bitcoin’s gossip efficiency. 
 Similarly, Berendae et al.~\cite{berendea2020fair} propose Hyperledger Fabric optimizations to reduce dissemination delay and improve fairness.
 While these optimizations focus on transaction propagation rather than DAS, they highlight the importance of optimizing dissemination networks, which aligns with \sysname's objectives.


%% file: sections/conclusion.tex
\vspacebeforesection
\section{Conclusion}
\vspaceaftersection
\label{sec:conclusion}

We presented \sysname, a practical approach to integrated data availability sampling (DAS) in the consensus workflow of Ethereum under the demanding Danksharding objectives.
By favoring direct and lightweight communications between nodes, builder-led blob data dissemination, and adaptive fetching mechanisms, \sysname allows large amounts of layer-2 data to propagate in the Ethereum network and be verified as available under the strict timing constraints imposed by Ethereum consensus.

This work opens interesting perspectives, among which is the design of \emph{adaptive} policies.
We presented and evaluated different fixed strategies for the builders and the nodes to follow.
However, the design could support automatic adaptation mechanisms that select or update parameters based, for example, on observed networking and fault ratio conditions.